\documentstyle[graphics,psfig]{article}
\newcommand{\be}{\begin{equation}}
\newcommand{\ee}{\end{equation}}
\newcommand{\bea}{\begin{eqnarray}}
\newcommand{\eea}{\end{eqnarray}}

\begin{document}
\title{Mode structure and ray dynamics of a parabolic dome microcavity}
\author{Jens U.~N{\"o}ckel\\
Max-Planck-Institut f{\"u}r Physik 
komplexer Systeme\\ N{\"o}thnitzer Str.~38, 01187 Dresden, Germany
\\[1ex]
Guillaume Bourdon, Eric Le Ru, Richard Adams,\\Isabelle Robert, Jean-Marie
Moison, and Izo Abram\\
Laboratoire CDP CNRS/CNET,\\
196 Avenue Henri Ravera, F-92220 Bagneux, France}
\date{published in Phys.~Rev.~E {\bf 62}, 8677 (2000)
}
\maketitle
\begin{abstract}
We consider the wave and ray dynamics of the electromagnetic field in a
parabolic dome microcavity.
The structure of the fundamental s-wave involves a main lobe 
in which the electromagnetic field is confined around the focal point 
in an effective volume of the order of a cubic wavelength, 
while the modes with finite angular momentum have a structure 
that avoids the focal area and have correspondingly larger effective volume.

The ray dynamics indicate that the fundamental s-wave is robust 
with respect to small geometrical deformations of the cavity, 
while the higher order modes are unstable giving rise to optical chaos.  
We discuss the incidence of these results on the modification 
of the spontaneous emission dynamics of an emitter placed 
in such a parabolic dome microcavity.
\end{abstract}
\tableofcontents

\section{Introduction}
The miniaturization of optoelectronic devices such as 
light emitting diodes or semiconductor lasers, 
is expected to lead to an improvement of their energy efficiency 
and to a lowering  of the lasing threshold. 
This tendency towards miniaturization has led to the exploration 
of optical microcavities whose dimensions 
are of the order of a few wavelengths \cite{Slusher}.
In such microcavities the extreme confinement of the electromagnetic field 
modifies the interaction of the active medium with the radiation field 
so that the process of spontaneous emission 
is altered both in its spatial and its dynamical characteristics.  
Spontaneous emission can thus be redirected, 
enhanced or inhibited in a way that may be exploited 
for the operation of light-emitting diodes or lasers.  
A modification of the characteristics of spontaneous emission, 
such as its directionality or the emission rate, 
has been shown for several microcavity designs such as 
for the traditional Fabry-Perot planar cavities \cite{Bjork}
and for disk-shaped \cite{disk} or spherical \cite{sphere} 
cavities displaying whispering gallery modes.

One of the key requirements for enhancing the dynamics 
of spontaneous emission and reducing the laser threshold is that 
the electromagnetic field at the site of the emitting dipole should be
enhanced inside the cavity with respect to its value in free space.  
A class of resonators for which this can be achieved very efficiently 
are confocal cavities: 
A few experiments with spherical confocal cavities \cite{Feld}, 
or semi-confocal microcavities \cite{Yamamoto} have been reported already, 
in which significant spontaneous emission modification 
or extremely low laser thresholds have been observed. 
Among the different designs of concave mirrors, {\em parabolic} 
mirrors have an important advantage in that their focal point
displays no astigmatism and is free from spherical aberrations. Basic 
geometric optic thus leads us to expect that 
double-parabolic confocal cavities or plano-parabolic semi-confocal 
cavities should display a strong enhancement 
of the electromagnetic field in the vicinity of the focal point, 
and a concomitant modification of the emission 
characteristics of an active medium placed there. 

\begin{figure}[tbp]
        \psfig{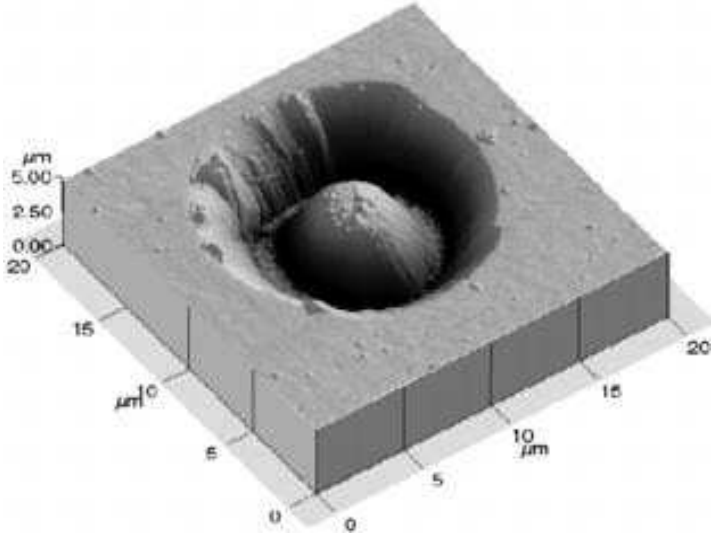}
        \caption{Atomic Force Microscope image of a ``hill" of 
diameter 7.2 $\mu$m and parabolic cross-section of height 1.8 $\mu$m,
etched on a GaAs substrate by a Focused Ion Beam apparatus.
When covered with gold it constitutes a concave parabolic mirror with its
focal point
inside the GaAs substrate.        \label{fig:afm-fib}
}
    \end{figure}

\begin{figure}[tbp]
        \psfig{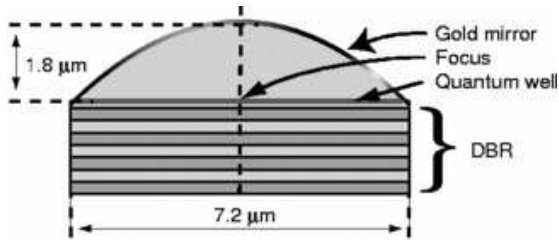}
        \caption{Schematic cross-section of a semiconfocal parabolic dome
cavity,
consisting of parabolic gold mirror and a planar Bragg mirror placed at the
focal plane of the parabola.
The cavity spacer is made of GaAs, and the light emitter is a quantum well
placed in the vicinity of the focal plane.        \label{fig:cav-schem}
}
    \end{figure}

This paper presents a theoretical analysis of microcavities formed by a 
parabolic mirror at or close to the confocal condition. The study is
motivated by experimental work in which such a system has in fact been 
fabricated. The experimental characterization of the modal structure and 
dynamics, being now in progress, will be given in a separate 
publication \cite{next}. Here, we briefly describe the experimental 
structure, in order to define the system for which our model 
calculations are intended. We have fabricated a 
semi-confocal plano-parabolic semiconductor microcavity 
(see Fig.\ \ref{fig:afm-fib})
by etching an appropriately-prepared GaAs wafer by a Focused Ion Beam
\cite{fib} to produce a ``hill" of cylindrical symmetry and parabolic 
vertical cross-section having a diameter of 7.2 $\mu$m 
and a height of 1.8 $\mu$m 
(corresponding to optical lengths of respectively 27 $\lambda $ 
and 6.75 $\lambda $ for a wavelength ({\it in vacuo}) of 960 nm)
which was subsequently covered with a thin metallic layer of gold.
This gold dome constituted thus a concave parabolic mirror 
with its focal point inside the GaAs substrate.
At the base of the parabolic hill, the wafer had a 6-period 
GaAs/AlAs Bragg mirror, closing the semi-confocal cavity (see Fig.\
\ref{fig:cav-schem}).
This cavity is expected to possess a mode in which the 
electric field is strongly enhanced in the vicinity of the focal point, 
so that a localized semiconductor emitter, 
such as a quantum box or quantum well, emitting at a wavelength near 960 nm,
will have its spontaneous emission greatly enhanced when placed there.
The use of a dielectric mirror with lower refractive index 
rather than a metallic mirror at the focal plane is important because 
it introduces a boundary condition that requires 
the tangential electric field to be maximal at the focal plane.
This condition cannot be fulfilled on a metallic mirror, 
on which the tangential electric field should vanish, 
producing thus a vanishing field at the focal point of the parabola.

In order to understand the operation of such a cavity 
and to assess its performance in modifying spontaneous emission,
in this paper, we examine first the modal structure of an ideal confocal
double-parabolic, or semi-confocal plano-parabolic microcavity.
We then investigate the stability of these modes with respect to geometric
deformations of the cavity that correspond to deviations from 
confocality; this condition is inevitably violated in a realistic cavity 
due to fabrication defects. The discussion of this case provides
the conceptual and theoretical background for the experimental 
analysis to be presented in a subsequent paper. 

The calculation of the modal structure of the parabolic dome microresonators
cannot be treated within the paraxial approximation of 
conventional \cite{siegman} 
resonator theory, because of the very large aperture displayed by the
parabola and because the cavity dimensions are comparable 
with the optical wavelength.  Extensions of the 
paraxial approximation to the highly convergent (or divergent) beams 
produced by parabolic mirrors are cumbersome even in macroscopic 
resonators \cite{laabs} where the optical axis is long compared with
the wavelength -- in microresonators, the latter breaks down as well. 
However, there are other approximate techniques which are well-suited 
to the problem we consider. As a valuable tool for simplifying the 
exact solution of Maxwell's equations for the cavity modes, we employ 
a short-wavelength approximation leading to simple WKB quantization 
conditions. The assumption that wavelengths are much shorter than the 
relevant cavity dimensions is common to both WKB and paraxial 
approximation, and it is therefore at first sight surprising that the 
WKB approach yields excellent quantitative agreement with the exact 
cavity spectrum even for the longest-wavelength modes of the 
parabolic cavity. We show how this arises by discussing in detail the 
structure of the classical ray dynamics in the resonator which makes 
the WKB approximation possible. As a result, we shall then also be able 
to assess the stability of the modal structure with regard to fabrication
imperfections, based on a ray analysis for parabolic cavities in cases 
where confocality is violated. To characterize the modes of 
the parabolic resonator, the internal caustic structure formed by the 
rays turns out to be of crucial importance. These considerations 
establish a connection between the the microcavity optics of the 
paraboloic dome and the field of quantum chaos: even minute 
deviations from confocality introduce chaos into the ray dynamics, and 
we have to address the significance of this effect for the relevant 
cavity modes.

The paper is organized as follows:
Section 2, introduces the mathematical model that describes confocal
parabolic cavities,
while Section 3 presents the wave equation for the electromagnetic field
in cylindrical and parabolic coordinates and discusses its exact vectorial
and scalar solutions.  
Section 4 presents the WKB approximation of the wave equations for the 
parabolic cavity,
an approach that will permit us in Section 6 to make the connection
with ray optics, while
Section 5 compares the numerical solutions of the wave equations
in the  parabolic microcavity with those of the WKB approximation.
Section 6 introduces the main concepts of ray optics applied to our
parabolic cavities with cylindrical symmetry, 
while Section 7 analyzes the stability of the ray trajectories in a
parabolic cavity in which there is a slight deviation from confocality.
Section 8 discusses the problem of the finite acceptance angle of Bragg mirrors,
a feature that limits the lifetime of modes in semiconfocal cavities
bounded by such mirrors.
Finally, Section 9 summarizes the results of this study and gives its
conclusions.

\section{The model}
We consider a model structure for an ideal semi-confocal cavity which is
bounded 
by a metallic concave parabolic mirror on one side and a planar 
dielectric mirror on the other side, placed at the focal plane of the
parabola.

In cylindrical coordinates ($\rho, z, \phi$) the parabolic mirror is given
by
\begin{equation} 
z = f- \frac{\rho^2}{4f}
\end{equation}
where $f$ is the focal distance of the parabola,
while the focal plane (and the planar mirror) corresponds to
\begin{equation}
z=0
\end{equation}
It is convenient to describe this cavity in parabolic coordinates $(\xi,
\eta, \phi)$, 
whose properties are summarized in the  Appendix. 
For reference, we reproduce here the transformation to cylindrical 
coordinates as given in Eq.\ (\ref{eq:revtransfapp}):
\begin{equation}\label{eq:revtransf}
\left \{ \begin{array}{l}
\rho = \sqrt{\xi\eta}\\
z = \frac12 (\xi - \eta)
\end{array} \right.
\end{equation}
To illustrate this coordinate system, we show in 
Fig.\ \ref{fig:doubleparab} (a) how the intersection of the 
coordinate surfaces defines a point A in the plane $z$ versus $\rho$. 
Also shown is the cavity shape itself:
    \begin{figure}[tbp]
        \psfig{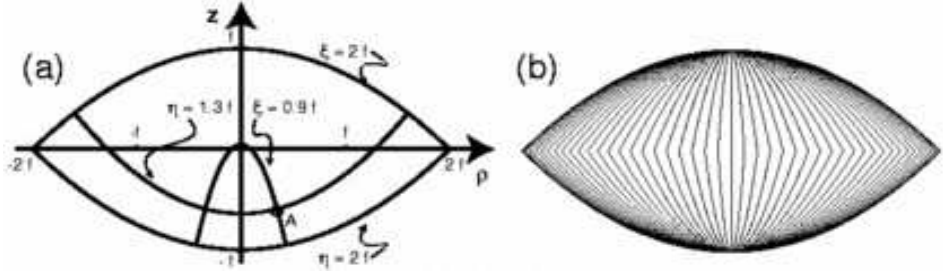}
        \caption{(a) Representation of parabolic cylinder coordinates in 
        the  $z$ - $\rho$ plane of a cylindrical coordinate system 
        ($z$ is the vertical axis).
        The third dimension is obtained by rotating the figure 
        around ${\hat z}$ by the angle $\phi$. The point A is 
        specified by $\xi=1.3\,f$, $\eta=0.9\,f$ and $\phi=0$. The 
        focus of all parabolas is at the origin.
	(b) By unfolding the parabolic dome into a double 
    paraboloid, the boundary conditions on the common focal plane 
    can be restated as simple parity requirement under reflection at 
    this plane ($z=0$). For TE modes, the electric field must be even 
    under this reflection. The unfolded cavity is shown in side view 
    with meridians which make $90^{\circ}$-corners at the focal plane. 
    The latter is also the equatorial plane of the cavity.    \label{fig:doubleparab}
}
    \end{figure}
the parabolic mirror corresponds to
\begin{equation}
\xi = 2\,f,
\end{equation}
and the planar dielectric mirror is at 
\begin{equation}
\xi = \eta.
\end{equation}

In an ideal cavity, the parabolic 
metallic mirror can be assumed to be lossless, displaying an 
amplitude reflectivity r = -1.  This produces a $\pi$ phase change 
upon reflection so that it corresponds to a boundary condition 
in which the tangential electric field vanishes.  
In parabolic coordinates this can be expressed as  
\begin{equation}
E_\eta(\xi = 2f)=0 ~~~~~~ E_\phi(\xi = 2f)=0 ~~~~~~B_\xi(\xi=2f)=0
\label{bdry-parab}
\end{equation}
Similarly, the planar dielectric mirror can be assumed to have a
reflectivity 
of r = +1, producing no phase change upon reflection so that 
the tangential magnetic field vanishes on the focal plane of the cavity.
In cylindrical coordinates, this can be expressed as
\begin{equation}\label{bdry-plane}
B_\rho (z=0) = B_\phi(z= 0)=0~~~\mbox{and}~~~ E_z(z=0)=0
\end{equation}
Alternatively, this implies that the tangential electric field 
is maximum on the focal plane and is symmetric under 
reflection of the whole cavity at the $z = 0 $ plane.
Thus, instead of considering this plane as an additional boundary with the 
properties (\ref{bdry-plane}), one can {\em unfold} the cavity 
across this plane by reflection, to obtain a confocal double 
paraboloid shown in Fig.\ \ref{fig:doubleparab} (b). 
This extended cavity requires only the metallic boundary conditions on its
parabolic 
walls, that is Eq.\ (\ref{bdry-parab}) and its equivalent in which $\xi $
and $\eta $ are interchanged.
It will support modes that can be either symmetric or 
antisymmetric under reflection at the focal plane. 
If we restrict ourselves to modes in which 
$E_\rho $ and $E_\phi $ are symmetric, this subset is identical to the modes
of the 
original dome with the conditions of Eqs.\ (\ref{bdry-parab}) 
and (\ref{bdry-plane}). 

The advantage of considering the unfolded 
cavity is that the focal plane as a physical boundary drops out of 
the discussion; this will considerably simplify the interpretation in 
terms of the ray picture later on.
Therefore, in the remainder of this paper, we can refer to 
Fig.\ \ref{fig:doubleparab} (b) as our model system.

\section{Wave equation }\label{sec:waveeq}
The electric field $\vec{E}$ obeys the vectorial wave equation
\begin{equation}
\nabla\times \nabla \times \vec{E} +
   \mu \epsilon \frac{\partial^2 \vec{E}}{\partial t^2}= 0
\label{vecwaveq}
\end{equation}
under the additional constraint that its divergence must vanish
\begin{equation}\label{eq:divfree}
\nabla \cdot \vec{E} = 0.
\end{equation}

The boundary conditions and the constraint of zero divergence 
imposed on the electromagnetic field in general will lead to a coupling 
between the various vectorial components of the electric and magnetic
fields. 
In simpler geometries such as cylinders, spheres or 
rectangular cavities, a suitable choice of polarizations reduces the 
problem to finding the eigensolutions of a scalar Helmholtz 
equation \cite{kerker}. However, in our case the three polarizations and the
intersecting 
parabolic surfaces forming the resonator cannot be labeled 
by the coordinate lines of a single orthogonal coordinate system, as 
is possible in the textbook systems mentioned. 
We discuss now the implications of this complication.

\subsection{Vector field components in cylindrical coordinates}
After combining Eqs.\ (\ref{vecwaveq}) and (\ref{eq:divfree}) to 
the wave equation,
\be\label{eq:wavestandard}
\nabla^2\vec{E}-\mu\epsilon
 \frac{\partial^2 \vec{E}}{\partial t^2}= 0,
\ee
we can take advantage of the cylindrical symmetry of the problem 
by expressing the wave equation for a time-harmonic 
electric field oscillating at frequency $\omega$
in cylindrical coordinates $(\rho, z, \phi)$, as
\begin{equation}
\left \{
\begin{array}{lc}
  \nabla^2 E_\rho - \frac{1}{\rho^2} E_\rho 
      - \frac{2}{\rho^2} \frac{\partial E_\phi}{\partial \phi} 
        + \mu \epsilon \omega^2 E_\rho = 0&(a)
\\ \\
  \nabla^2 E_\phi - \frac{1}{\rho^2} E_\rho
      + \frac{2}{\rho^2} \frac{\partial E_\rho}{\partial \phi} 
        + \mu \epsilon \omega^2 E_\phi = 0&(b)
\\ \\
  \nabla^2 E_z + \mu \epsilon \omega^2 E_z = 0&(c)
\end{array}
\right .
\label{cylindeq}
\end{equation}
We note that the wave equation couples the radial and angular 
components of the electric field ($E_\rho$ and $E_\phi$),
while the equation for the axial component $E_z$ is scalar. One can 
achieve a further simplification in this system of 
equations as follows:

The rotational symmetry around the $z$ axis permits us to assume 
a $\phi$-dependence of all components of the field of the form 
\be\label{eq:separansatz}
Q(\rho , z) \cdot e^{im\phi}.
\ee
With this ansatz, Eqs.~(\ref{cylindeq} a,b) can be written as
\begin{equation}\label{eq:transversefield}
\left\{ 
\begin{array}{lc}
\rho ^{2} \left[ \nabla^{2} +k^{2} \right] E_{\rho } 
-E_{\rho } =2im E _{\phi } &(a) \\ \\
\rho ^{2} \left[ \nabla^{2} +k^{2} \right] E _{\phi } -E
_{\phi } =-2im E _{\rho } &(b)
\end{array}
\right. 
\end{equation}
where $k = \sqrt{\mu \epsilon} \omega$ is the wavenumber inside the 
parabolic dome. If the azimuthal quantum number $m=0$, this 
reduces to two identical equations. If, on the other hand, $m\neq 0$, 
we can form a suitable linear combination of $E_{\rho}$ 
and $E_{\phi}$ which decouples these two equations. Naively setting 
$E_{\phi}=0$ would not achieve this goal because it forces both 
field components to vanish. 

The proper linear combination in which to decouple this system of 
differential equations is obtained with the definition
\begin{equation}
\left\{ 
\begin{array}{l}
E _{\rho } \equiv\frac{i}{\sqrt{2} } (E _{+} -E _{-} ) \\ \\
E _{\phi } \equiv\frac{1}{\sqrt{2} } (E _{+} +E _{-} ).
\end{array}
\right. 
\end{equation}
Then $E _{\pm } $ are the solutions of the equations
\begin{equation}\label{eq:circwave}
\rho ^{2} \left[ \nabla^{2} +k^{2} \right] E_{\pm} 
= (1 \pm 2m) E_{\pm}
\end{equation}
This definition again makes use of the azimuthal symmetry of the 
resonator, which implies that the 
circular polarizations 
$\hat{\sigma}_\pm = \mp \frac{i}{\sqrt{2}} \left ( \hat{\rho} 
\pm i \hat{\phi} \right )$
are decoupled in the cylindrical wave equation. 
In this way, we have therefore formally decoupled the original system 
of equations Eq.\ (\ref{cylindeq}) for the vector field components.
In the special case $m=0$, case $E_{+}$ and 
$E_{-}$ will moreover be linearly dependent because their respective 
equations again coincide. 

However, this decoupling of polarizations in the wave equation does 
not reduce the problem to a truly scalar one because the field 
components are still coupled by the boundary conditions and by the condition
of zero divergence.
On the ``top" parabolic mirror, the conditions (\ref{bdry-parab}) 
in terms of the cylindrical components of the electric field now read:
\begin{equation}\label{eq:newboundarycond}
\begin{array}{ll}
\mbox{at} ~~~ \xi=2f & \left \{
\begin{array}{l}
i\sqrt{f} (E_+-E_-) + \sqrt{\eta} E_z = 0 \\
\\
E_++E_- = 0 \\
\\
\frac{\partial}{\partial \eta} (E_++E_- )= 0.
\end{array}
\right .
\end{array}
\end{equation}
The first line expresses the condition $E_{\eta}=0$, the second and 
third lines represent $E_{\phi}=0$ and $B_{\xi}=0$, respectively. 
On the ``bottom" parabolic mirror the boundary conditions are the same as
in Eq.\ (\ref{eq:newboundarycond}) with $\xi $ and $\eta $ interchanged.

\subsection{The absence of longitudinal electromagnetic modes}
Unfortunately, the set of boundary conditions Eq.\
(\ref{eq:newboundarycond})
is not yet a complete list of constraints that we have to satisfy. 
An additional requirement is that the field at every point in the resonator 
has to have zero divergence, which in parabolic coordinates reads
\bea\label{eq:zerodiv}
E_+ - E_- + m (E_++E_-) + 
\frac{2\xi\eta}{\xi+\eta} \cdot
 \left ( \frac{\partial}{\partial \xi} +
    \frac{\partial}{\partial \eta} \right ) (E_+-E_-) &&\nonumber\\
 +   \frac{2i \sqrt{2\xi\eta}}{\xi+\eta} \cdot
 \left ( \xi \frac{\partial}{\partial \xi} +
    \eta \frac{\partial}{\partial \eta} \right ) E_z
&=& 0
\eea
This assumption already entered the derivation of the 
system of wave equation, Eq.\ (\ref{eq:wavestandard}), from the original 
Maxwell equations in the form of Eq.\ (\ref{vecwaveq}). However, this 
does not guarantee that all solutions of  Eq.\ (\ref{eq:wavestandard}) 
or Eq.\ (\ref{cylindeq}) 
satisfy Eq.\ (\ref{eq:zerodiv}). The latter is just the 
well-known statement that the electromagnetic field is purely {\em 
transverse}, ruling out longitudinal modes: the 
transverse electric field $\vec{E}_{\perp}$ is 
related to the curl of the magnetic field by the Maxwell equation
\be 
\nabla\times\vec{B}=\frac{1}{c}\frac{\partial\vec{B}}{\partial 
t}=ik\vec{E_{\perp}},
\ee
and hence satisfies $\nabla\cdot\vec{E}_{\perp}=0$; the longitudinal 
field $\vec{E}_{\parallel}$, which can 
be written as the gradient of a potential $\Phi$, is responsible for 
violations of Eq.\ (\ref{eq:zerodiv}). 

In view of the constraints imposed by the boundary conditions
(\ref{eq:newboundarycond}) and by the zero divergence condition
(\ref{eq:zerodiv}) it is not possible (except for the case $m = 0 $,
as we shall see later) to set one (or two) of the vector components
to zero without setting the full electric field identically to zero,
and thus it is not possible to reduce in a rigorous manner
the vector problem into a scalar one.  

The problem can in principle be solved by converting  
Eq.\ (\ref{eq:zerodiv}) from a condition in the cavity {\em volume} 
to a {\em boundary condition} which can then be treated on the same 
footing as Eq.\ (\ref{eq:newboundarycond}). One way of achieving this 
\cite{balianbloch} is by noting that if
\be
\vec{E}_{0}\equiv
\vec{E}_{\perp}+\vec{E}_{\parallel}
\ee
fulfills Eq.\ (\ref{eq:wavestandard}), then so does
\be
\vec{E}\equiv k^2\nabla\times\nabla\times\vec{E}_{0}=
 k^2\nabla\times\nabla\times\vec{E}_{\perp}.
\ee
The latter is automatically divergence-free. In order for $\vec{E}$ 
to satisfy the boundary condition $\vec{E}_{t}=0$, we 
require for $\vec{E}_{0}$ 
\be\label{eq:transversefield1}
\left(\vec{E}_{0}\right)_{t}=0\quad{\rm and}
\nabla\cdot\vec{E}_{0}=0
\ee
on the surface. Then one has indeed
\bea
\vec{E}_{t}&=&
 k^2\left(\nabla\times\nabla\times\vec{E}_{0}\right)_{t}\nonumber\\
&=&k^2\,\left(\vec{E}_{0}+\nabla(\nabla\cdot
\vec{E}_{0})\right)_{t}=k^2\,\left(\vec{E}_{0}
\right)_{t}=0.
\eea
on the boundary. The problem is therefore reduced to finding the 
auxiliary field $\vec{E}_{0}$ and then deducing the transverse field 
from Eq.\ (\ref{eq:transversefield1}). This leads to a system of three 
second-order differential equations for each vector component 
of $\vec{E}_{0}$, given by Eq.\ (\ref{cylindeq}), 
all of which are coupled by boundary conditions that are, however, 
quite complex.

The next step is then to write the field components as linear 
combinations of independent general solutions of Eq.\ (\ref{cylindeq}) 
and determine the unknown coefficients in that expansion from the 
matching conditions at the boundary. The solution proceeds in an 
analogous but much less tedious way if we neglect the additional 
divergence condition. The important simplification is that we are 
then able to consider the $E_+$ and $E_-$ components of the electric 
field independently, by setting all except for one component to zero. 
The boundary conditions (\ref{eq:newboundarycond}) are then decoupled 
as well. More 
precisely, it will be shown that the wave equations are then not only 
scalar but also {\em separable}, i.e., reducible to the solution of 
ordinary differential equations.

We therefore would like to neglect the coupling 
that results from the condition of zero divergence, provided that this 
can be justified in the context of the present study. There are various 
reasons why this approximation will provide us with useful results.
Foremost, it will turn out below that the most important modes we 
find in this way in fact conspire to satisfy Eq.\ (\ref{eq:zerodiv}) 
{\em a posteriori}, cf.\ Section \ref{sec:fundawave}: 
the modes that provide the best confinement of the field in a tightly 
focused region around the focal point are the ones with $m=0$. For 
these, the different vector components decouple rigorously and the 
scalar program is exact. These $m=0$ modes are particularly 
significant because they provide the best 
confinement of the field in a tight focal volume. This is the 
paramount aim of the experimental dome structure. 

In addition to this exact result, the more transparent simplified 
problem allows us to evaluate the stability of the stationary states 
of the field in the
parabolic cavity with respect to deviations from the 
confocality condition -- a deformation that can readily occur in the 
course fabrication. This will be addressed with the help of the ray 
picture in Section \ref{sec:chaos}, and the ray trajectories 
themselves are independent of whether a vectorial or scalar field is 
considered. Since the exact nature of the deformation is 
unknown, it is necessary to make model assumptions and parametrize 
the deformation in some way. Although the range of possible behaviors 
explored within our model can be argued to be generic, we lose at that 
point the ability to predict accuratey all the individual modes of the 
specific sample. The error incurred by this fundamental uncertainty 
about the precise boundary shape is larger than the error made 
by adopting the simplified boundary conditions, and hence the latter 
are warranted on physical grounds. 

The consistency of these arguments is proven in Section \ref{sec:chaos} 
where we find that the only modes which {\em can} in fact be reliably 
predicted for a large range of possible deformations (because they 
are structurally stable against the emergence of chaos) are the ones 
with low $m$ (or angular momentum in the classical picture), 
concentrated strongly near the $z$-axis. For these modes one can set 
approximately $m\approx 0$, $E_{+}=E_{-}$ and $E_{z}=0$ so that 
Eq.\ (\ref{eq:zerodiv}) becomes valid. 

\subsection{The wave equation in parabolic coordinates}
Having discussed the boundary conditions, we now provide the 
solutions to Eq.\ (\ref{eq:circwave}). 
In order to find a system of general solutions to the formally 
scalar differential equations, Eqs.\ (\ref{cylindeq} c) or 
(\ref{eq:circwave}), we express the scalar Laplacian appearing there 
in parabolic coordinates ($\xi, \eta, \phi$), leading to the form 
\begin{equation}\label{eq:kummerdeqn0}
\frac{4}{\xi+\eta} \cdot
 \left [
    \frac{\partial}{\partial \xi}
       \left ( \xi \cdot \frac{\partial Q}{\partial \xi} \right )
   + \frac{\partial}{\partial \eta}
       \left ( \eta \cdot \frac{\partial Q}{\partial \eta} \right )
 \right ]
+ k^2 Q
= \frac{n^2}{\xi\eta} \cdot Q
\end{equation}
where
\begin{equation}\label{eq:kummercases}
\left \{
\begin{array}{lll}
n = m + 1 & for & Q = E_+\\
n = m - 1 & for & Q = E_-\\
n = m & for & Q = E_z
\end{array}
\right .
\end{equation}
Here we have used the fact that the derivative $\partial^2/\partial\phi^2$ 
appearing in the Laplacian $\nabla^2$ pulls down a factor $-m^2$ due 
to the ansatz Eq.\ (\ref{eq:separansatz}). Although the righthand side 
is the analog of the centrifugal barrier in cylindrical problems, it 
thus depends not on angular momentum $m$ directly but on a modified 
azimuthal mode number $n$. This occurs due to the additional $\phi$ 
derivatives introduced when we transformed the vector field components to 
cylindrical coordinates in Eq.\ (\ref{cylindeq}).

At this point we introduce the approximation of discarding the 
divergence condition so that we merely have to consider the boundary 
conditions (\ref{eq:newboundarycond}) with one and only one of 
the three field components nonzero. Then, 
Eq.\ (\ref{eq:kummerdeqn0}) is separable in $\eta$ and $\xi$. 
We shall return to the details of the solution 
procedure in Section \ref{sec:scalarparab}; for now it is sufficient 
to give the result: 
Denoting the separation constant by $\beta$, the solution can be written 
in the form 
\be\label{eq:kummerseparate}
Q = F(k,\beta, \xi) \cdot F(k,-\beta, \eta)
\ee
where  $F(k,\beta, \xi)$ obeys
\begin{equation}\label{eq:kummerdeqn}
\xi F'' + F' + 
\left (- \frac{n^2/4}{\xi} + \frac{k^2}{4} \xi + \beta \right ) F = 0
\end{equation}
The functions $F(k,\beta, \xi)$ and $F(k,-\beta, \eta)$
appearing here are solutions of 
this differential equation with the same $k$ and $n$, but with 
sign-reversed $\beta$, and hence their functional dependence on $\xi$ 
and $\eta$ will be different unless $\beta=0$.  
Without loss of generality, we can assume $n$ 
to be nonnegative, because it appears in the above equation only as 
$n^2$. 
The solutions that do not diverge at $\xi = 0$ are of the form
\begin{equation}\label{eq:kummersoln}
F(k,\beta, \xi) = e^{ik\xi/2}\xi^{\frac{n}{2}}
              M \left ( \frac{n+1}{2}-\frac{i\beta}{k}; n+1 ; -ik\xi \right
)
\end{equation}
where $M(a,b,z)$ is Kummer's confluent hypergeometric function. 
The function $F$ as written here is in fact real, because of the 
Kummer transformation \cite{abramovitz}
\be
M(\frac{b}{2}-a,b,-z)=e^{-z}\,M(\frac{b}{2}+a,b,z),
\ee
where we set $a=i\beta/k$, $b=n+1$ and $z=ik\xi$. Appplying the 
theorem then yields $F(k,\beta, \xi)=F(k,\beta, \xi)^*$. 

The separation constant $\beta$ and the wavenumber $k$ at which to 
find the mode are still unknowns of the problem that have to be 
determined from the boundary conditions. 
The first constraint we can write down is 
\be\label{eq:ekline1}
F(k,\beta,2\,f)\equiv 0
\ee
to enforce vanishing tangential field on the parabolic surface. 
In the two-dimensional plane spanned by the unknowns 
$\beta$ and $k$, this single equation defines a set of curves. 
The boundary condition on the focal plane requires that $E_{\pm}$ be 
symmetric under reflection, i.e., invariant under 
$\xi\leftrightarrow\eta$. For $E_{z}$, on the other hand, one needs 
odd parity. In order to construct such solutions with a well-defined 
parity, we have to form linear combinations 
\be\label{eq:superprod}
E=F(k,\beta,\xi)\,F(k,-\beta,\eta)
\pm F(k,-\beta,\xi)\,F(k,\beta,\eta),
\ee
where in addition 
\be\label{eq:ekline2}
F(k,-\beta,2\,f)\equiv 0. 
\ee
The set of curves parametrized by this constraint will intersect the 
curves defined by Eq.\ (\ref{eq:ekline1}) at certain isolated 
{\em points} in the $\beta$ - $k$ plane. By finding these 
intersection points, we determine the quantized values of $\beta$ and 
$k$ corresponding to solutions of Eq.\ (\ref{eq:kummerdeqn0}) which 
satisfy the boundary conditions. It is not clear at this stage of the 
discussion how many intersections there are, or even how the curves 
defined by each equation separately will look. 
Before we analyze the different branches of these equations and 
identify their intersections based on asymptotic methods in Section
\ref{sec:scalarparab}, it is useful to discuss in more detail the
consistency of the fields thus obtained.

\subsection{Behavior at the focal point}\label{sec:focalpoint}
The main experimental purpose of the cavity is 
to concentrate the field near the focus as much as possible. Since 
one always has $E_{z}=0$ there, it remains to discuss the behavior 
of $E_{\pm}$ in the focal region. 
Because of the ``angular-momentum-barrier'' on the righthand side of 
Eq.\ (\ref{eq:kummerdeqn0}), the solutions $F$ given in 
Eq.\ (\ref{eq:kummersoln}) attain a factor $\xi^{n/2}$ which 
suppresses the field near the origin $\xi=0$ when $n\neq 0$. The 
Kummer function itself goes to $M=1$ at $\xi=0$, so that the only way of 
getting a nonvanishing field at the origin is to set $n=0$
in Eq.\ (\ref{eq:kummerdeqn}). This means that the angular momentum quantum 
number must in fact satisfy $m=1$ for $E_{-}$ or $m=-1$ for $E_{+}$ 
according to Eq.\ (\ref{eq:kummercases}). But this leads to a contradiction:
if the field is nonzero at the origin, then because of the azimuthal 
factor $\exp(\pm i\phi)$ one faces a singularity at $\xi=\eta=0$ in which 
the field is indeterminate. Therefore, there is {\em no possibility} to 
obtain a nonzero field precisely at the focus of the cavity. 

For $m=\pm 1$ there are still solutions of Eq.\ (\ref{eq:kummerdeqn0}), 
but they must involve solutions of Eq.\ (\ref{eq:kummerdeqn0}) in 
suitable linear combinations such as to yield a vanishing field at 
$\xi=\eta=0$. 
We have the freedom to linearly combine eigenstates of the wave 
equation at the same wavenumber $k$ (yielding a stationary state with 
monochromatic time dependence). First we use the real-valued 
solutions in Eq.\ (\ref{eq:kummersoln}) to form a 
superposition of the type Eq.\ (\ref{eq:superprod}) with a plus sign. 
Despite its symmetry it 
can also be made to vanish at $\xi=\eta=0$, if one or both of the 
functions $F(k,\beta,\xi)$ and $F(k,-\beta,\eta)$ are zero at the origin. 

Although we can hence find 
solutions for arbitrary $m$ with a tangential electric field that 
is symmetric under reflection at the focal plane, we can only attempt 
to concentrate the field {\em near} the focus, always with a node at 
the focal point, dictated for $m\neq 0$ by the phase singularity at 
the origin. For $m=0$, there is the residual angular momentum barrier 
due to $n\neq 0$, and thus even in this simple case -- contrary to our 
expectation from quantum mechanical analogues -- the ``s-wave'' 
solutions have vanishing field at the focus, 
as a consequence of the vector nature of the field.

\subsection{Particular case: the fundamental 
s-wave}\label{sec:fundawave}
The case $m=0$ can be discussed in more detail because it permits simple 
analytical expressions for the wave solution, if we specialize further 
to $\beta=0$. In this case, the solutions in Eq.\ (\ref{eq:kummersoln}) 
simplify to 
\begin{equation}\label{eq:kummersolnsimpl}
F(k,0, \xi) \propto
    {\rm I}(\frac{n}{2},\frac{i}{2}\,k\xi)\,i^{n/2},
\end{equation}
dropping prefactors that are absorbed in the normalization. Here, 
${\rm I}$ is the modified Bessel function. 

As was already noted below Eq.\ (\ref{eq:circwave}), 
$E_{\pm}$ linearly dependent in the special case $m=0$, 
so that we can in particular 
choose $E_{+}=E_{-}$. Then $E_\rho=0$ and $E_+=E_-=E_\phi/\sqrt{2}$.
We thus obtain the TE field by setting 
\be
E_{\phi}=Q=F(k,0, \xi)\,F(k,0, \eta)
\ee
as in Eq.\ (\ref{eq:kummerseparate}). This already satisfies the 
condition of symmetry with respect to the focal plane, without having 
to form a superposition of the type (\ref{eq:superprod}). Moreover, 
it satisfies the condition of vanishing divergence, as can be checked 
with Eq.\ (\ref{eq:zerodiv}). 

With $n=1$ (for $E_{+}$ at $m=0$), 
Eq.\ (\ref{eq:kummersolnsimpl}) can be rewritten to obtain 
\begin{equation}\label{eq:m0field}
\vec{E} = \left \{\begin{array}{l}
  E_\xi = 0\\
  E_\eta = 0 \\
  E_\phi = E_0 \cdot \frac{1}{k\sqrt{\xi \eta}} \cdot \sin(k\xi/2)
\sin(k\eta/2)
\end{array}\right.
\end{equation}
and
\begin{equation}
\vec{B} = \left \{ \begin{array}{l}
   B_\xi = -i E_0 \cdot \frac{\sqrt{\mu \epsilon}}{k} \cdot
\sqrt{\frac{1}{\xi+\eta}} \cdot
             \frac{1}{\sqrt{\xi}} \sin(k\xi/2) \cos(k \eta/2)
\\
   B_\eta = +i E_0 \cdot \frac{\sqrt{\mu \epsilon}}{k} \cdot
\sqrt{\frac{1}{\xi+\eta}} \cdot
             \frac{1}{\sqrt{\eta}} \cos(k\xi/2) \sin(k \eta/2)
\\  
   B_\phi = 0
\end{array}\right.
\end{equation}

The resonance condition is obtained from the boundary condition 
Eq.\ (\ref{bdry-parab}) on the parabolic dome at $\xi = 2f$ as
\begin{equation}\label{eq:resoconditio} 
k_N = N \frac{\pi}{f}, \quad N=1,2,\ldots
\end{equation}

A relatively simple visualization of these modes can be obtained by
expressing the electric and magnetic fields in
cylindrical and spherical coordinates which are more familiar.  It 
can be verified using the relations between these coordinates to the 
parabolic variables that Eq.\ (\ref{eq:m0field}) then takes the form 
\begin{equation}
\vec{E} = \left \{\begin{array}{l}
  E_\rho = 0\\
  E_z = 0 \\
  E_\phi = E_0 \cdot \frac{1}{k\rho} \cdot \left ( \cos(kz)- \cos(kr) 
  \right ),
\end{array}\right.
\end{equation}
and the corresponding magnetic field is
\begin{equation}
\vec{B} = \left \{ \begin{array}{l}
B_\rho = i E_0 \cdot \frac{\sqrt{\mu \epsilon}}{k} \cdot \frac{1}{r \rho} (
z \sin(kr)-r \sin(kz) )\\
B_z = i E_0 \cdot \frac{\sqrt{\mu \epsilon}}{k} \frac{\sin(kr)}{r} \\
B_\phi = 0
\end{array} \right.
\end{equation}
By splitting the various terms appearing here into two contributions, 
the electromagnetic field can then be considered as the superposition 
of two fields:

The first field is polarized along $E_\phi$ and $B_\rho$ and can be 
expressed in cylindrical
coordinates as
\begin{equation}\label{eq:firstfield}
E_\phi^{(1)} = \frac{E_0}{k_N \rho} \cos(k_N z) ~~~\mbox{and}~~~
B_\rho^{(1)} = i \sqrt{\mu \epsilon} \frac{E_0}{k_N \rho} \sin(k_N z)
\end{equation}
The second field is polarized in spherical coordinates ($r$, $\phi$, 
$\theta$) along 
the directions of the azimuthal and polar angles, $\phi$ and $\theta$, 
according to 
\begin{equation}\label{eq:secondfield}
E_\phi^{(2)} = \frac{E_0}{\sin\theta} \frac{\cos(k_N r)}{k_N r}
~~~\mbox{and}~~~
B_\theta^{(2)} = i \sqrt{\mu \epsilon} 
\frac{E_0}{\sin\theta}\frac{\sin(k_N r)}{k_N r} 
\end{equation}
Here, we have used the substitution $\rho=r\,\sin\theta$ in the 
denominators. 

\begin{figure}[tbp]
    \centering
    \psfig{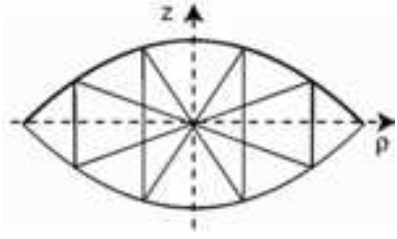}
    \caption{Cross-sectional view of the unfolded cavity with two 
    closed, bowtie-shaped ray paths going through the common focus of 
    the bounding parabolas. Families of such rays can be thought of 
    as constituents of the s-wave in Eq.\ (\ref{eq:m0field}).     \label{fig:n0bowtie}
}
\end{figure}

The first field, Eq.\ (\ref{eq:firstfield}), 
corresponds to cylindrical standing waves with a phase
variation along the $z$ direction, while the second 
field, Eq.\ (\ref{eq:secondfield}), corresponds to spherical
standing waves with a phase variation along the radial direction.
This configuration is reminiscent of what is expected from a simple 
geometrical optics argument in which a ray bundle 
emerging from the focal point can propagate outwards
as a spherical wave, upon reflection on the parabola it gets 
converted into a cylindrical wave,
which in turn can counter-propagate back to the focal point
after being reflected on the planar mirror and a second time on the 
parabola. In the unfolded double-paraboloid, the ray trajectories 
are of the type shown in Fig.\ \ref{fig:n0bowtie}.
It should be noted that these two partial waves are not physical when taken 
individually,
because in both cases the electric field diverges along the axis of the
parabola.
The divergence, however, cancels out when the superposition of the two
partial 
waves is considered. We have so far only drawn this ray interpretation 
from a particular decomposition of the exact solution; the 
question is how arbitrary that decomposition is and what we can learn 
from it. This will be pursued in the following section. The actual 
intensity distribution of these $n=1$ states in the cavity will be plotted
in 
Section \ref{sec:modevolume} where we can compare their spatial 
patterns with those obtained for larger $n$, in order to justify our 
claim that the s-wave modes provide the best focussing. 

\section{Finding the modes within the short-wavelength approximation}
\label{sec:scalarparab}
Having seen that even the long-wavenlength s-wave in our cavity
can be interpreted as standing waves arising from counterpropagating 
ray bundles and their accompanying wavefronts, we now turn to a 
more quantitative eikonal analysis. 
Such an analysis can provide accurate starting points 
for a numerical search of the exact wave solutions, which are 
determined by finding intersection points between the families of
curves (\ref{eq:ekline1}) and (\ref{eq:ekline2}) in the plane 
of $\beta$ vs. $k$. Such semiclassical considerations based on the 
short-wavelength 
approximation are an important first step because there are, as we shall 
see, infinitely 
many intesections between the sets of curves determining the exact 
solutions, and one desires a 
means of finding them in a systematic way, labeling them by ``quantum 
numbers'', giving the number of nodes in the field along the 
coordinate lines for $\xi$ and $\eta$. Beyond this very 
practical use of the short-wavelength limit, we also want to 
establish a physical understanding of the resonator modes that allows 
us to predict how they depend on changes in the cavity shape. This 
aspect of the ray picture will be expounded in the last section. 

\subsection{WKB approximation and effective potential}
The equation to be solved is Eq.\ (\ref{eq:kummerdeqn}), an ordinary 
second-order differential equation, where the angular momentum $m$ 
enters as a parameter trough the constant $n$. 
We are looking for solutions $F(\xi)$
which satisfy the boundary condition $F(2\,f)=0$ and are not singular  
at $\xi=0$. The standard short-wavelength approach to be employed here 
is the WKB approximation\cite{mathews}. After division by $\xi$, 
Eq.\ (\ref{eq:kummerdeqn}) takes the form 
\be\label{eq:newxideqn}
\frac{d^2f}{d \xi^2}+\frac{1}{\xi}
\frac{df}{d \xi}+\frac{1}{4}\left(k^2+\frac{4\beta}{\xi}-
\frac{n^2}{\xi^2}\right)\,f(\xi)=0.
\ee
For the subsequent analysis it is
convenient to introduce a dimensionless coordinate
\be\label{eq:uresc}
u=\sqrt{k\,\xi}.
\ee
Dividing Eq.\ (\ref{eq:newxideqn}) by $k^2$ and defining a rescaled 
separation constant
\be
Z\equiv \frac{4\,\beta}{k},
\ee
the following equation is obtained:
\be\label{eq:udeqn}
-\frac{d^2f}{du^2}-\frac{1}{u}\frac{df}{du}
+\frac{1}{4}\left(
\frac{n^2}{u^2}-u^2\right)\,f(u)=Z\,f(u).
\ee

This has a form similar to the one-dimensional Schr{\"o}dinger equation 
of quantum mechanics, except for the first $u$-derivative which makes 
the kinetic energy operator non-selfadjoint. This term appears in 
the radial equation of cylindrically symmetric problems but does not 
affect the applicability of the WKB approximation \cite{substitution}.

\begin{figure}[tbp]
    \centering
    \psfig{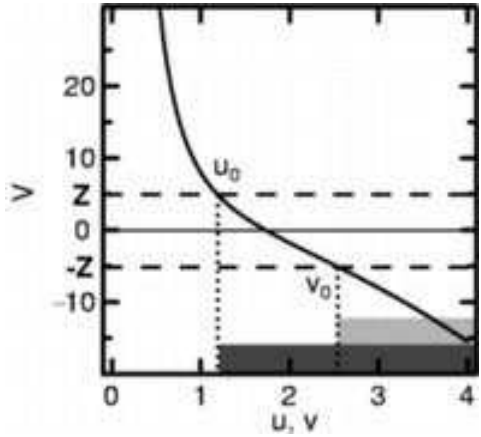}
    \caption{Solid curve: the effective potential 
    $V(u)$ for $n=3$, showing the 
    classical turning points $u_{0}$ ($v_{0}$) where $Z$ ($-Z$)
    intersects the effective potential $V$, cf. the dashed lines. 
    The ranges of 
    classically allowed motion for the two degrees of freedom $u$ and 
    $v$ with energies $Z$ and $-Z$ are indicated by the shaded bars 
    (dark for $u$, light for $v$). The outer turning points at 
    $u,\,v=\sqrt{2kf}$ act as a hard wall whose position depends on 
    $k$.    \label{fig:classpotent}
}
\end{figure}

The WKB solution requires us to find the classical turning points in 
the potential appearing in this equation, with $Z$ playing the role 
of the total energy. This effective potential,
\be\label{eq:potential}
V(u)=\frac{1}{4}\left(\frac{n^2}{u^2}-u^2\right),
\ee
is a superposition of an inverted parabola and the centrifugal 
potential determined by $n$, giving rise to the solid line in 
Fig.\ \ref{fig:classpotent}.
Using this together with the ansatz
\be\label{eq:wkbansatz}
f(u)\approx\frac{1}{p(u)}\,e^{i\int p(u)\,du},~~~p(u)=\sqrt{Z-V(u)},
\ee
the approximate solutions are found by imposing the boundary 
conditions at the turning points. 

There is only one possible turning point corresponding to the closest 
approach to the origin $u=0$, which is given by
\be\label{eq:uturnpoint}
V(u_{0})=Z\Rightarrow u_{0}=\sqrt{\sqrt{n^2+\frac{Z^2}{4}}-\frac{Z}{2}}.
\ee
If $Z>0$ and $n=0$, then no inner turning point exists. This inner 
turning point, in classical mechanics, is the point where the momentum 
in the $x$-direction smoothly goes through zero as it changes sign, and
hence 
the probability per unit time of finding the particle becomes 
infinite. In the ray dynamics, this phenomenon gives rise to a {\em 
caustic}. This will be discussed further in Section 
\ref{sec:caustics}. 

The outer turning point of this classical picture is determined by the 
Dirichlet boundary condition at the parabolic mirror, which in the new 
coordinate is located at 
\be
\xi_{1}=2\,f\Rightarrow u_{1}=\sqrt{2\,kf}.
\ee
It is the boundary condition $f(u_{1})=f(\sqrt{2\,kf})=0$ in which 
the short-wavelength condition is contained: we assume that at the 
outer boundary the wavefunction has the WKB form, 
Eq.\ (\ref{eq:wkbansatz}), which requires that
the dimensionless {\em size parameter} satisfies 
\be
x\equiv 2\,kf\gg 1,
\ee
i.e. this additional boundary is far away from the classical turning 
point $u_{0}$ of the effective potential. 
All steps discussed above for $f(\xi)$ apply analogously to the 
variable $\eta$ appearing in the product ansatz $Q$, 
Eq.\ (\ref{eq:kummerseparate}), if we reverse the sign of $Z$ 
everywhere and replace $u$ by the variable 
\be\label{eq:vdef}
v=\sqrt{k\eta}.
\ee
Then the inner turning point $v_{0}$ for this second degree of freedom is 
obtained as
\be\label{eq:vturnpoint}
V(v_{0})=-Z\Rightarrow v_{0}=\sqrt{\sqrt{n^2+\frac{Z^2}{4}}+\frac{Z}{2}}.
\ee
The values of $u_{0}$ and $v_{0}$ determine the distance of closest 
approach to the $z$ axis

\subsection{Quantization conditions}
Under this condition, the semiclassical determination of the 
eigenfrequencies proceeds by applying the Bohr-Sommerfeld quantization 
to the action integral for one period of the motion in 
the effective
potential. One round trip consists of the path from $u_{0}$ to $u_{1}$ 
and back to $u_{0}$. The quantized action is therefore
\be
J(Z,x;n,\nu)\equiv 2\,\int\limits_{u_{0}}^{u_{1}}\sqrt{Z-V(u)}\,du\equiv
2\pi\,
\left(\nu+\frac{3}{4}\right).
\ee
The integer $\nu=0,\,1\,\ldots$ is the number of nodes of the wavefunction 
in the 
potential, and the constant $3/4$ takes into account the phase shifts 
of $\pi$ and $\pi/2$, at the outer and inner turning 
points, respectively.  In other words, the above quantization 
condition is an approximate way of writing the phase-shift 
requirements that hold at boundaries and caustics, using the 
approximation that the wave propagation itself is described by a 
wavefront whose phase advance in $x$ is given by the function $J$. 

The result of the integration is found to be
\bea
J(Z,x;n,\nu)&=&
\sqrt{x^2+Z\,x-n^2}\nonumber\\
&&+\frac{Z}{2}\,\ln\frac{\sqrt{x^2+Z\,x-n^2}+x+Z/2}
{\sqrt{n^2+Z^2/4}}\nonumber\\
&&-n\,\left(\arcsin\frac{Z\,x-2 n^2}{x\sqrt{Z^2+4n^2}}
+\frac{\pi}{2}\right)\nonumber\\
&=&2\,\pi\,\left(\nu+\frac{3}{4}\right).
\label{eq:actionquant}
\eea
This is an equation for the two unknowns $Z$ and $x$, i.e. for the 
rescaled separation constant and size parameter. It is therefore 
analogous to Eq.\ (\ref{eq:ekline1}). The integer $\nu$
uniquely labels all the allowed solutions $f_{\nu}$ of 
Eq.\ (\ref{eq:udeqn}). This is an important difference to 
Eq.\ (\ref{eq:ekline1}): there, the function $F$ in fact has 
infinitely many {\em branches} that satisfy the equation, which are 
however not labeled explicitly. The great advantage of 
Eq.\ (\ref{eq:actionquant}) is that these branches are explicitly 
enumerated by $\nu$, so that fixing this index selects exactly one 
curve in the $Z$-$x$ plane instead of an infinte family. 

As in  Eq.\ (\ref{eq:kummerseparate}), the field consists of products 
of the form 
\be
Q = f_{\nu}(x,Z,u) \,f_{\mu}(x,-Z,v)
\ee
with $v$ defined as in Eq.\ (\ref{eq:vdef}).
The two function $f_{\nu}$ and $f_{\mu}$ have their analog in the 
exact solutions $F$ of Eq.\ (\ref{eq:kummerdeqn}), corresponding to 
the branches of $F$ labeled by $\nu$ and $\mu$, respectively. 

\begin{figure}[tbp]
    \centering
    \psfig{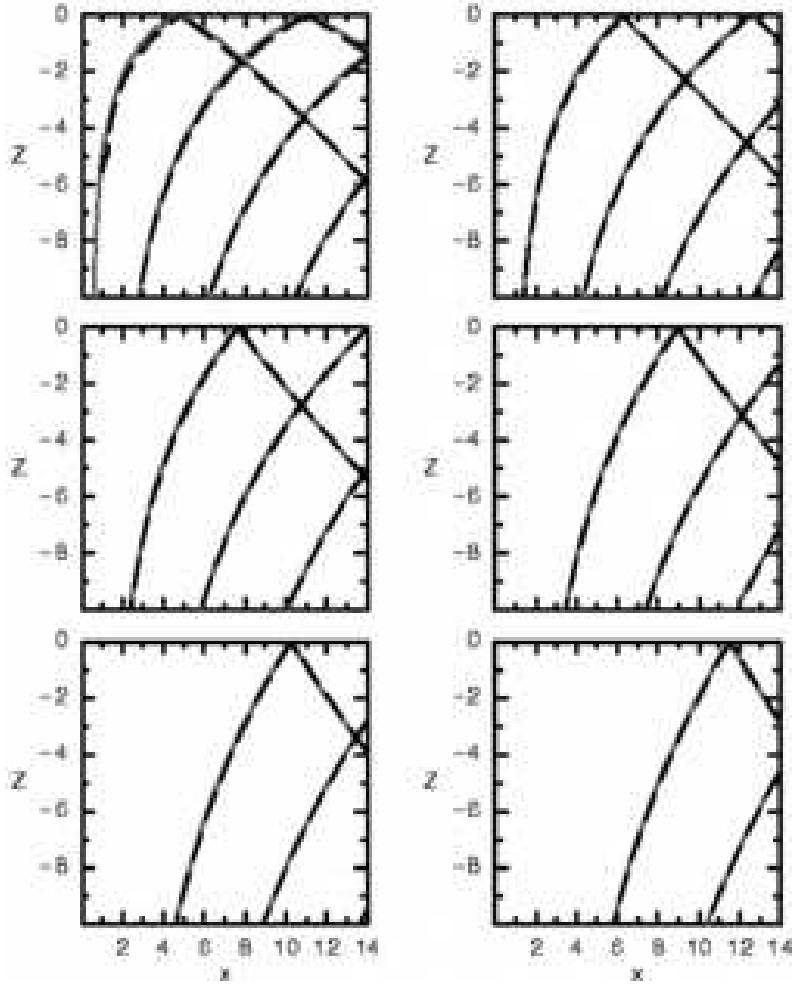}
    \caption{For the first six values of $n$, the graphical solution 
    of the simultaneous equations (\ref{eq:actionquant}) and
    (\ref{eq:actionquant2}) can be read off from the intersections of 
    the red curves. The dashed black curves show the analogous graphical 
    solution of Eqs.\ (\ref{eq:ekline1}) and (\ref{eq:ekline2}). The 
    exact and semiclassical curves are almost indistinguishable 
    (except for $n=0$), 
    attesting to the striking accuracy of the former even at the 
    smallest possible size parameters $x$. All plots can be continued 
    to $Z>0$ by reflecting at the axis $Z=0$. The WKB curves 
    with positive slope belong to Eq.\ (\ref{eq:actionquant2}), the 
    falling lines are created by Eq.\ (\ref{eq:actionquant}). They are 
    labeled starting from the leftmost by $\mu,\,\nu=0,\,1,\,2\ldots$. 
        \label{fig:graphsolution}
}
\end{figure}

We then form combinations the form of Eq.\ (\ref{eq:superprod}) to 
enforce the required symmetry with respect to the focal plane.
The semiclassical WKB quantization for the function $f_{\mu}(x,-Z,v)$ 
provides a second equation of the form (\ref{eq:actionquant}),
\be\label{eq:actionquant2}
J(-Z,x;n,\mu)=2\,\pi\,\left(\mu+\frac{3}{4}\right).
\ee
These two quantization conditions play the same role as 
Eqs.\ (\ref{eq:ekline1}) and (\ref{eq:ekline2}): the intersections of 
the curves parametrized by them determine the quantized values of $Z$ 
and $x$. However, the WKB method affords a great simplification: 
by fixing the branches $\nu$ and $\mu$, the 
intersection of the two resulting curves is uniquely determined.
To illustrate this situation, we 
show in Fig.\ \ref{fig:graphsolution} how the lines defined by the 
above two equations traverse the $Z$-$x$ plane. Only a small portion
of this plane is shown, emphasizing the behavior of the semiclassical 
results at small $x$ where their accuracy should be at a minimum. 
Comparison with the exact families of curves shows, however, that the 
WKB results are excellent even in this long-wavelength limit. 
Note that by symmetry, intersections occurring at $Z=0$ are always 
between curves with the same branch index $\nu=\mu$. All curves shift 
to larger $x$ with increasing $n$ because of the larger centrifugal 
barrier, pushing the classically allowed regions of the effective 
potential in Fig.\ \ref{fig:classpotent} outward. 

\section{Exact solution for the modes and their field distribution}
Once the allowed combinations of $Z$ and $x$ -- or equivalently 
$\beta$ and $k$ -- at which the boundary 
conditions are satisfied have been found, the problem of finding the 
modes is solved. For example, we can now plot the intensity 
distribution of each mode by using the quantized values of 
$\beta$ and $k$ in Eq.\ (\ref{eq:kummersoln}) and forming the proper 
linear combinations of the form Eq.\ (\ref{eq:superprod}). 

\subsection{Mode profiles}
This will now be carried out for the lowest-lying modes as obtained 
from the intersections in Fig.\ \ref{fig:graphsolution}. 
Any given value of $n$ can have a different meaning for the intensity 
distribution in the azimuthal direction, depending on which case in 
Eq.\ (\ref{eq:kummercases}) we choose to consider: $n=m\pm 1$ for the 
modes. Since the azimuthal field variation is trivial, 
$\propto\exp(im\phi)$, we wish to 
restrict our attention to the mode profile in the plane spanned by 
$\rho$ and $z$ in cylindrical coordinates. 
The variable governing this property is $n$, not $m$. 
Therefore, $n$ is used here to classify the mode profiles. 

\begin{figure}[tbp]
    \centering
    \psfig{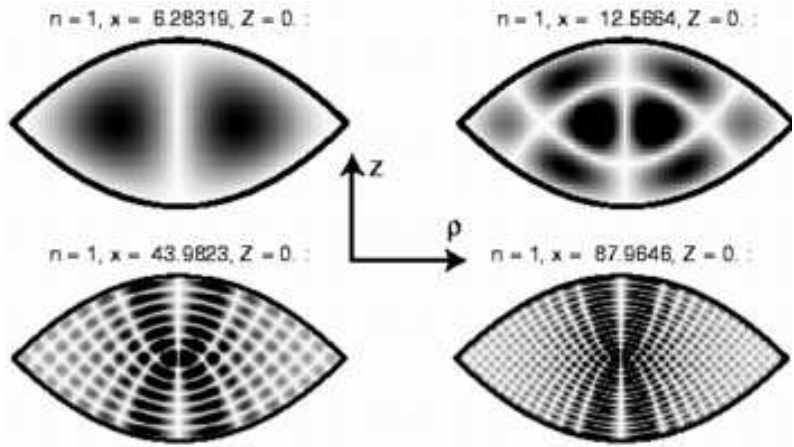}
    \caption{Four states with $n=1$. Grayscale indicates the 
    magnitude of the electric field for the modes ($E_{\pm}$)
    of the confocal double paraboloid, highest fields shown in black.
    The vertical axis is $z$, the horizontal axis the axial distance $\rho$.
    Increasing $x$ means shorter wavelength and hence more nodal 
    lines (white). The size parameter is quantized according to 
    Eq.\ (\ref{eq:sizeparaquantn0}) with $N=1,\,2$ in the top row, 
    and $N=7,\,14$ at the bottom.    \label{fig:symprofilesn1}
}
\end{figure}

\begin{figure}[tbp]
    \centering
    \psfig{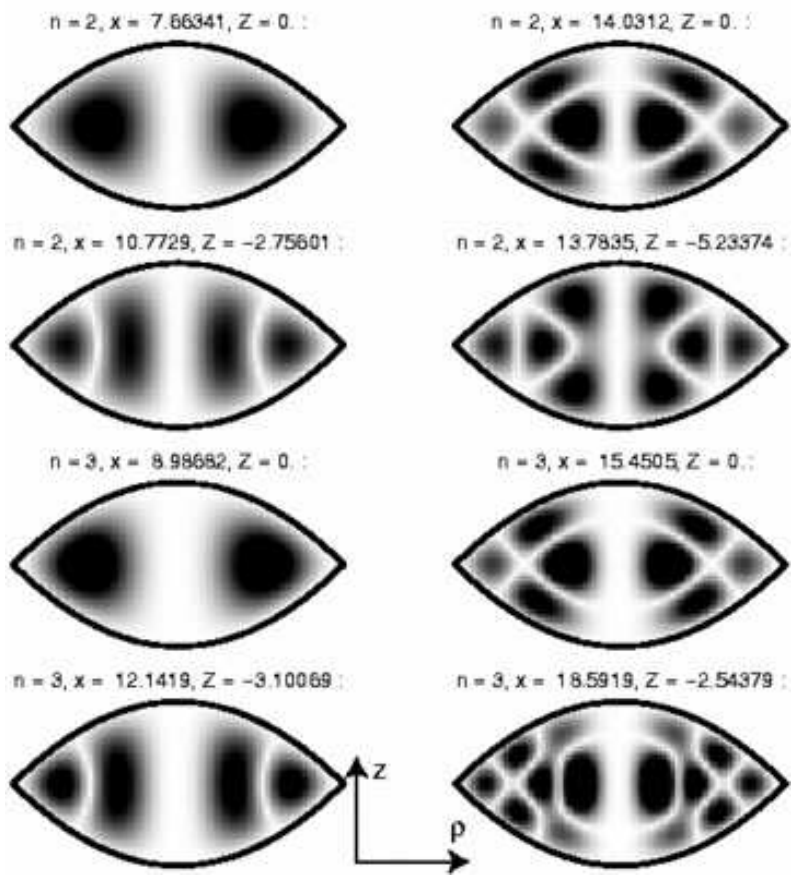}
    \caption{The modes shown here do not follow the simple law of 
    Eq.\ (\ref{eq:sizeparaquantn0}) but were obtained numerically. 
    With increasing centrifugal barrier, labeled by $n$, the forbidden 
    region around the $z$-axis grows outward. 
        \label{fig:symprofiles}
}
\end{figure}

As has been done in the previous sections, we shall take the focal 
plane to be the symmetry plane of a double paraboloid, and plot the 
wave fields in this unfolded cavity. This is done in view of the 
subsequent discussion, where we shall establish the connection between 
these modes and the ray dynamics. Some wave plots are shown in 
Figs.\ \ref{fig:symprofilesn1} and \ref{fig:symprofiles}. 
Note that the case $n=0$ does not appear among the solutions listed 
here because it corresponds to wave 
fields that do not vanish on the $z$-axis and hence are 
irreconcilable with the finite angular momentum $m=\pm 1$, as discussed 
in section \ref{sec:focalpoint}). 

If we look at only the left column of Figs.\ \ref{fig:symprofilesn1} and 
\ref{fig:symprofiles}, it is apparent that all states with $Z=0$ look 
similar, as do all states with $Z\neq 0$. A similar observation can be 
made in the right columns of the figures. Comparison to the 
intersecting lines in the graphical solution, 
Fig.\ \ref{fig:graphsolution}, shows that states with the same nodal 
pattern indeed result from the crossing of the same pair of lines 
-- labeled by the same $\mu$ and $\nu$, only for different $n$ which 
pushes the intersecting lines to higher $x$. 
However, the interpretation of $\mu$, $\nu$ as the number of nodes in 
the parabolic coordinate directions cannot be carried through in all 
of the plots. We will return to this problem in Section 
\ref{sec:desymmetrized}.

In order to achieve the best possible concentration of fields near the 
focus, the most promising candidates are the modes with $n=1$. Among 
these, the patterns shown in Fig.\ \ref{fig:symprofilesn1} indicate 
that the states at $Z=0$ in turn show the highest intensity near the 
focal plane. These are precisely the fundamental s-waves we discussed 
in Section \ref{sec:fundawave}, with the wavenumbers quantized 
according to Eq.\ (\ref{eq:resoconditio}), which for the size 
parameter reads
\be\label{eq:sizeparaquantn0}
x_{N}=2\,k_{n}f=2\,\pi\,N.
\ee
This is an exact result which can be compared to the 
WKB quantization condition in Eq.\ (\ref{eq:actionquant}) with 
$Z=0$, $n=1$. The latter actually has a more complicated form,
\be
\sqrt{x^2-1}+\arcsin\frac{1}{x}=2\pi\,(\nu+1),
\ee
but to second order in the small quantity $1/x$ this is identical 
to Eq.\ (\ref{eq:sizeparaquantn0}) with $N=\nu+1$. This confirms the 
observation made in Fig.\ \ref{fig:graphsolution} that the numerical 
agreement between exact and semiclassical solutions is good even for 
small quantum numbers.

\subsection{Focussing and the effective mode 
volume}\label{sec:modevolume}
In order to evaluate the field enhancement that is achieved in the
fundamental
TE modes discussed in Section \ref{sec:fundawave} and shown in 
Fig.\ \ref{fig:symprofilesn1}, 
it is necessary to examine the distribution of the electromagnetic energy
in that mode.
The energy in a parabolic cavity of focal length $f$ is
\begin{eqnarray}
U & =& \frac14 \int^{\xi=2f}_{\xi=0} \int^{\eta=2f}_{\eta=0}
\int^{\phi=2\pi}_{\phi=0} 
           \nonumber \\
 & &	\left ( \frac{\epsilon}{2} 
	(\left| E_{\xi} \right|^2 +\left| E_{\eta} \right|^2+ \left|
E_{\phi} \right|^2 ) +
	 \frac{1}{2\mu} 
	(\left| B_{\xi} \right|^2 +\left| B_{\eta} \right|^2 +\left|
B_{\phi} \right|^2 )
	\right ) \nonumber \\
 & &  \frac{\xi+\eta}{4} \, d\xi \, d\eta \, d\phi
\end{eqnarray}
which gives, for the fundamental (s-wave) TE modes, Eq.\ (\ref{eq:m0field}),
\begin{equation}\label{eq:energy}
U = \epsilon E_0^2 \frac{\pi f}{4 k^2} 
\int^{kf}_0 \frac{\sin^2(x)}{x} \, dx
\end{equation}
where the value of the integral can be evaluated numerically.

For the experimentally realized 
cavity described in Section 1, $kf = 14\,\pi$ so that the value of
the integral is $2.527$. The intensity distribution for this mode is 
shown at the bottom right of Fig.\ \ref{fig:symprofilesn1} (note $x=2\,kf$).
To examine the energy distribution in the cavity,
we can evaluate the energy that is contained at each lobe of the standing
wave 
of parabolic wavefront that corresponds to the mode.
We note then that the first lobe, corresponding to a parabolic wavefront of
focal length $f_1 = \lambda/2$,  contains $48$\% of the total energy; 
to see this, replace the integration limit in Eq.\ (\ref{eq:energy}) 
by $\pi$. This lobe occupies a physical volume of $V_{0}\equiv\pi 
\lambda^3/4$ whereas the volume of the overall cavity, $V=2\pi\,f^3$, 
is $2744$ times larger. 

This underscores the very large confinement of the field that occurs in the 
vicinity of the focal point and points to the possibility of observing 
a very large enhancement of spontaneous emission into this mode. The 
fraction of the total energy contained in the first lobe of course 
reaches $100$\% if the smallest possible cavity with $k\,f=\pi$ is 
considered. However, the size achieved in our present sample already 
approaches the optimal conditions if one takes into account that 
enhancement of spontaneous emission requires 
not only a small effective mode volume but most of all a small local 
density of states \cite{yamamotoslusher}. The {\em average} 
density of modes in an 
arbitrarily-shaped electromagnetic resonator of volume $V$ is a fundamental 
quantity in the theory of blackbody radiation and was derived by H.~Weyl 
\cite{weyl}:
\be\label{eq:weyl3d}
\rho_{\rm Weyl}(k)\approx \frac{k^2}{\pi^2}\,V.
\ee
Note that this can also be written in the physically intuitive form 
\be\label{eq:weyl3da}
\rho_{\rm Weyl}(k)\approx 
\frac{2}{3}\pi^2\,\frac{d}{dk}\left(\frac{V}{V_{0}}\right),
\ee
indicating that the number of modes in the interval $dk$ is 
proportional to the number of additional volume quanta $V_{0}$ that 
fit into the given volume $V$ when $k$ increases to $k+dk$. The {\em 
local} density of states in the focal volume $V_{0}$ can therefore 
be interpreted to be the same as the total density of states in a 
small cavity of volume $V=V_{0}$. This, in turn, is roughly the 
effective mode volume for the fundamental s-wave in our structure. 
>From this we conclude that the spontaneous emission enhancement should 
be close to 
the maximum possible value even though our cavity is not of the 
minimum possible size. This is one of the central advantages we were 
looking for in the parabolic cavity design. In this discussion we have 
assumed for simplicity that the Q-factor of the modes under 
consideration is fixed, independent of size and quantum numbers. 
This severe simplification will be removed in Section 
\ref{sec:rayescape}.

In the higher order modes with $m>0$, 
the centrifugal barrier prevents the field from approaching the focal point.
This implies that these modes will have a larger effective volume 
and, correspondingly, a smaller enhancement of the spontaneous emission
rate.  
An added difficulty concerning the higher order modes arises from the
limited experimental control over the exact cavity shape.
As discussed in Section \ref{sec:chaos},
small deformations of the cavity (modeled as deviations from confocality),
result in chaos, leading to a loss of constraints on the possible regions 
of phase space which can be explored. This 
further increases the effective volume of these modes.
The enhanced spontaneous emission into the fundamental s-wave
implies that this mode will also exhibit a large gain and, correspondingly,
a low lasing threshold. The preliminary conclusion of this 
section is therefore that a mode 
with low angular momentum and small $Z$ (or $\beta$) 
will be the dominant mode in a laser of parabolic
geometry. 

\subsection{Caustic structure in the wave solutions}
\label{sec:desymmetrized}
In order to arrive at the solutions shown in Fig.\ \ref{fig:symprofiles}, 
we started from the semiclassical (short-wavelength) approximation 
and then refined the quantized $Z$ and $k$ further by applying the 
exact modal conditions. However, the question arises how the 
quantum numbers $\mu$ and $\nu$ which label the semiclassical 
solutions can be visualized in Fig.\ \ref{fig:symprofiles}. The answer is 
that the symmetrization procedure obscures this identification. What 
happens can be understood if we ignore the parity requirement and 
plot the wave fields in the simple product form of 
Eq.\ (\ref{eq:kummerseparate}). 

\begin{figure}[tbp]
    \centering
    \psfig{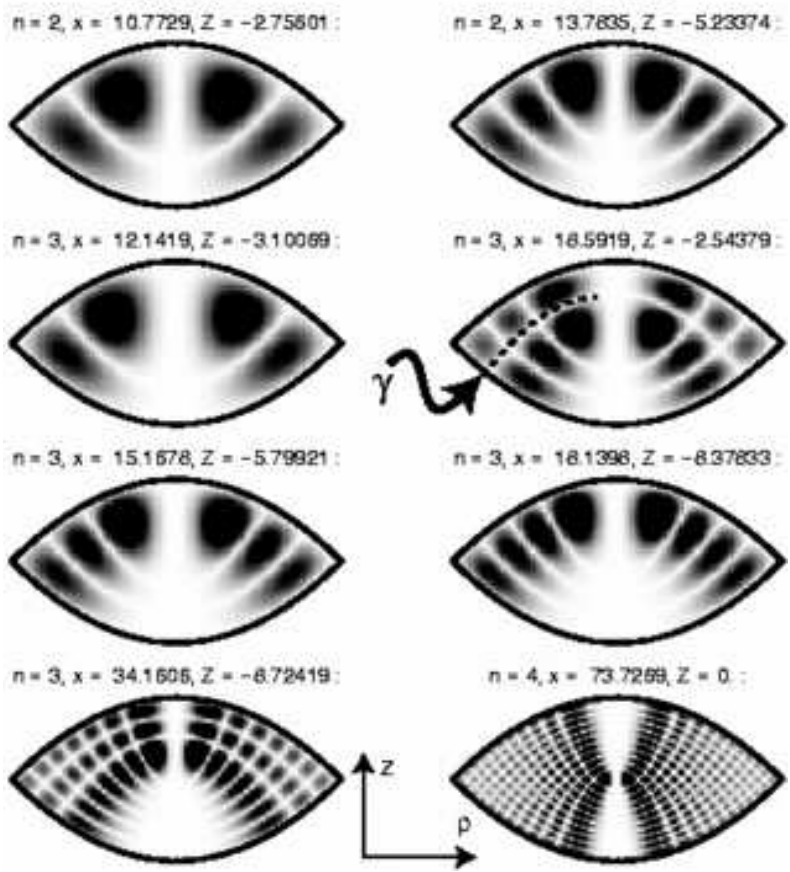}
    \caption{Mode intensities as in Fig.\ \ref{fig:symprofiles}, 
    but without performing the symmetrization prescribed by the focal 
    plane boundary condition.  The WKB quantum numbers 
    $\mu$ (and $\nu$) can be read off by counting the number of 
    wavefunction nodes parallel (and perpendicular) to the reference 
    line $\gamma$. Modes in the first three rows correspond to the 
    symmetrized versions of Fig.\ \ref{fig:symprofiles}. In order to 
    illustrate the approach to the short-wavelength limit, additional 
    modes are shown for which the formation of caustics is apparent in 
    the high-intensity ridges (black) bordering the classically forbidden 
    regions (white).
        \label{fig:unsymprofiles}
}
\end{figure}

The symmetrization performed according to Eq.\ (\ref{eq:superprod}) 
with $A=B$ introduces no change whatsoever if the separation constant 
is $\beta=0$. Therefore, the intensity profiles of all modes with 
$Z=0$ in Fig.\ \ref{fig:symprofiles} are the same before and after 
symmetrization. However, the wave patterns acquire a qualitatively 
different and simpler form if we 
desymmetrize the remaining states. This is shown in 
Fig.\ \ref{fig:unsymprofiles}. 
The nodal patterns now appear in a regular fashion along the 
coordinate lines for $u$ and $v$ (or $\xi$ and $\eta$), and their 
number along these lines is uniquely determined by $\mu$ and $\nu$. 

By symmetrizing a state such as the one shown for 
$n=3$, $k=18.59$ and $Z=-2.54$ in 
Fig.\ \ref{fig:unsymprofiles}, the field shown in the desymmetrized 
plot is added to its reflection at the focal plane, thus allowing 
some nodal lines to be ``filled in'', as seen in the corresponding 
state at the bottom right of Fig.\ \ref{fig:symprofiles}. The 
desymmetrized waves in Fig.\ \ref{fig:unsymprofiles} exhibit nodal 
lines precisely along lines of $\eta=const$ or $\xi=const$. In 
addition to the simple nodal structure, we also observe a clear 
segregation between regions of negligible intensity and regions of 
oscillatory field, with dividing lines between them that become more 
and more pronounced as the size parameter $x=2\,k\,f$ increases. These 
are the caustics, which in fact accumulate an increasing amount of 
intensity as the short-wavelength limit is approached. The caustics 
follow parabolic coordinate lines as well, as is apparent from the 
last row of Fig.\ \ref{fig:unsymprofiles}. The field at $n=3$, 
$x=34.16$, $Z=-8.72$ is bounded from below by a broad inverted 
parabola, and excluded from the $z$-axis by a steep upright 
parabola. The intersection of both parabolas forms the caustic. In 
the mode at $n=4$, $x=73.73$, $Z=0$, both the upright and inverted 
bounding parabolas are symmetric as we expect for $Z=0$. 

\section{Caustic structure in the ray picture}\label{sec:caustics}
In this section we will elaborate on the relation between mode 
structure and ray dynamics, as a basis on which we can predict the 
effect of shape perturbations on the mode structure.
The caustic patterns revealed in the last section 
by the decomposition into the product states as in 
Eq.\ (\ref{eq:kummerseparate}) is a direct consequence of the 
classical turning points in the effective potential $V$, 
Eq.\ (\ref{eq:potential}), for the motion along the $\xi$ and $\eta$ 
directions. The distinction between classically allowed and forbidden 
regions gives rise to the regions of oscillatory and vanishing fields 
in Fig.\ \ref{fig:unsymprofiles}. The effective potential has, so far, 
been discussed only as an auxiliary concept that proved 
convenient in the WKB treatment; its relation to the behavior of the 
rays of geometric optics is, however, well-known. For the sake of 
a self-contained presentation, we convey here the idea behind the 
general eikonal theory by showing how to derive ray equations from 
the one-dimensional separated wave equations, 
Eq.\ (\ref{eq:kummerdeqn}). The argument is non-standard in the sense 
that Eq.\ (\ref{eq:kummerdeqn}) is based on the full vectorial wave 
equation (i.e.\ with polarization), and we therefore shall find that 
for a given angular momentum $m$, slightly different ray trajectories 
have to be considered depending on polarization. This is because the 
quantity entering Eq.\ (\ref{eq:kummerdeqn}) is $n$, not $m$. 

\subsection{Ray equations from the WKB approximation}
Inserting the WKB ansatz, Eq.\ (\ref{eq:wkbansatz}), into the wave 
equation for the separated variables, Eq.\ (\ref{eq:kummerdeqn}), 
one finds to leading semiclassical order that $p$ 
must satisfy the equation
\be
p_{u}^2+V(u)=Z,~~~\mbox{similarly}~p_{v}^2+V(v)=-Z.
\ee
We can interpret this as the Hamiltonians of two decoupled linear 
systems, and add them to obtain the Hamiltonian for the combined 
system, 
\be
{\tilde H}=p_{u}^2+p_{v}^2+V(u)+V(v)
\ee
The trajectories we are looking for then satisfy the equation 
${\tilde H}(p_{u},p_{v},u,v)=Z-Z=0$, or written out:
\be
p_{u}^2+p_{v}^2+\frac{1}{4}\,
\left(\frac{n^2}{u^2}+\frac{n^2}{v^2}-u^2 -v^2\right)=0.
\ee
If we divide this by $(u^2+v^2)$, the result is
\be
\frac{p_{u}^2+p_{v}^2}{u^2+v^2}+\frac{1}{4}\,
\left(\frac{n^2}{u^2v^2}-1\right)=0.
\ee
this can also be interpreted as arising from a {\em new} Hamiltonian
\be
H\equiv\frac{p_{u}^2+p_{v}^2}{u^2+v^2}+\frac{1}{4}\,
\frac{n^2}{u^2v^2}
\ee
by requiring 
\be\label{eq:newhamliton}
H(p_{u},p_{v},u,v)=\frac{1}{4}.
\ee
The Hamiltonian in this form is analogous to the wave equation in 
parabolic coordinates, Eq.\ (\ref{eq:kummerdeqn0}),
where the Laplacian is divided by the same 
scale factor $(u^2+v^2)$ that accompanies the conjugate momenta 
here. One can now use Hamilton's equation of motion to replace momenta 
by ``velocities'', the definition being
\be
{\dot u}=\frac{\partial H}{\partial p_{u}},~
{\dot v}=\frac{\partial H}{\partial p_{v}}.
\ee
This leads to the substitution 
\be
p_{u}=\frac{1}{2}\,(u^2+v^2)\,{\dot u},~
p_{v}=\frac{1}{2}\,(u^2+v^2)\,{\dot v},
\ee
which brings Eq.\ (\ref{eq:newhamliton}) into the form
\be
(u^2+v^2)\,({\dot u}^2+{\dot v}^2)+\frac{n^2}{u^2v^2}=1.
\ee
Reverting to cylinder coordinates, the above equation becomes 
\be\label{eq:raycyl1}
k^2\,({\dot \rho}^2+{\dot z}^2)+\frac{n^2}{k^2\,\rho}=1.
\ee
Here we used the definitions of the coordinates in 
Eqs.\ (\ref{eq:revtransf}) and (\ref{eq:uresc}). 
To examine what this equation has to do with the ray 
dynamics, we take the ray-picture point of view now. 

\subsection{Geometric optics in cylindrical coordinates}
If we consider the three-dimensional motion of rays in a double 
paraboloid of the shape in Fig.\ \ref{fig:doubleparab} (b), their 
propagation between reflections at the parabolic walls will of course 
follow straight lines, and hence there is no place for any 
coordinate-dependent potential $V$. However, in order to compare the 
ray dynamics to wavefunction plots in the $z$-$\rho$ plane as shown in 
Fig.\ \ref{fig:unsymprofiles}, we must project the ray motion onto 
this plane as well. In the wave analysis, this projection was 
achieved by using the {\em rotational symmetry} of the cavity 
around the $z$ axis to eliminate the azimuthal coordinate $\phi$ 
from the problem in favor of the angular momentum quantum number $m$.

In ray optics, we can do the same: 
rays can be classified by an angular momentum $L_z$ because 
of the axial symmetry. To see this, 
we first define $L_z$. A ray trajectory is a curve 
consisting of straight line segments between each reflection. If we 
parametrize this curve as ${\bf r}(l)$, where $l$ is the path length 
along the ray from some arbitrary starting point, then $|{\dot{\bf 
r}}(l)|=1$. Here and in the following, the dot represents the 
differentiation with respect to arc length, $d/dl$.
In cylinder coordinates $\rho,\,\phi,\,z$, we can 
decompose this as
\be\label{eq:unitvector}
{\dot{\bf r}}={\dot \rho}\,{\bf e}_{\rho}+{\dot z}\,{\bf 
e}_z+r\,{\dot\phi}\,{\bf e}_{\phi}.
\ee
Between any two reflections, this is a constant unit vector in the 
direction of the ray. With this, 
the equation for a straight line segment can be written in general as
\be
{\bf r}\times{\dot{\bf r}}\equiv{\bf L},\label{lineeqn}
\ee
where ${\bf L}$ is a constant analogous to the angular momentum of 
classical mechanics. 

Because of the rotational symmetry around the $z$-axis, the azimuthal 
unit vector ${\bf e}_{\phi}$ at the point of reflection is always 
tangent to the surface. Therefore, a reflection does not change the 
component of ${\dot{\bf r}}$ along ${\bf e}_{\phi}$, so that 
$\rho\,{\dot\phi}$ is continuous. Since the ray curve is itself 
continuous everywhere, so is $\rho(l)$. Hence the quantity
\be\label{eq:angmom}
L_z\equiv \rho^2\,{\dot\phi}
\ee
is also continuous at each reflection. But this is just the 
$z$-component of ${\bf L}$ in Eq.\ (\ref{lineeqn}), as can be 
verified by performing the cross product there. Thus, $L_z$ is a 
constant between reflections, which together with its overall 
continuity implies that it is a conserved quantity for the whole ray 
trajectory. 

Using Eq.\ (\ref{eq:angmom}), the fact that $\dot{\bf r}$ is a 
unit vector, Eq.\ (\ref{eq:unitvector}), can be recast as
\be\label{eneryconseqn}
\dot{\rho}^2+\dot{z}^2 + \frac{L_z^2}{\rho^2} = 1.
\ee
>From the ray approach we have thus obtained an equation almost 
identical to Eq.\ (\ref{eq:raycyl1}). We only have to re-define the 
path length variable $l$ to make it dimensionless, by introducing
\be
s=k\,l,
\ee
to obtain for the derivatives 
\be
\frac{d\rho}{dl}=k\frac{d\rho}{ds},
\ee
and interpret furthermore 
\be\label{eq:lzsemicl}
L_{z}=\frac{n}{k}.
\ee
Then Eqs.\ (\ref{eq:raycyl1}) and (\ref{eneryconseqn}) become 
identical, if we interpret the dot in Eq.\ (\ref{eq:raycyl1}) to 
mean $d/ds$. The scale factor of the ``time'' variable parametrizing 
our trajectories is irrelevant for the shape of the paths, so that we 
can conclude that 
{\em the ray picture introduced here is equivalent to the motion 
described by the WKB effective potential, with the important 
identification of Eq.\ (\ref{eq:lzsemicl})}.

Besides Eq.\ (\ref{eneryconseqn}), 
the only other equation that is needed to completely determine any 
ray trajectory from its initial conditions is the law of specular
reflection, which can be formulated with the help of the outward
normal unit vector $\bf u$ at the reflection point as
\be\label{speculareqn}
{\dot{\bf r}}_{reflected} = 
{\dot{\bf r}} - 2\,{\bf u}\,({\bf u}\cdot{\dot{\bf r}}).
\ee
This corresponds to a reversal of the normal component of 
${\dot{\bf r}}$. Here we can see explicitly that reflections do not
affect the component of ${\dot{\bf r}}$ in the direction of 
${\bf e}_{\phi}$, since the normal $\bf u$ has no ${\bf
e}_{\phi}$-component as a consequence of the axial symmetry. 

This latter fact also means that we can simply drop the 
${\bf e}_{\phi}$-component from Eq.\ (\ref{speculareqn})
altogether. Therefore, we now define the two-component vectors in the 
$z$ - $\rho$ plane by dropping the ${\bf e}_{\phi}$-components from the
corresponding three-component vectors. Thus, ${\dot{\bf r}}$ becomes
\be\label{2dveleqn}
{\bf v}\equiv {\dot \rho}\,{\bf e}_{\rho}+{\dot z}\,{\bf e}_z
\equiv\left(\begin{array}{c}{\dot\rho}\cr{\dot
z}\end{array}\right),
\ee
and similarly
\be
{\bf u}= u_{\rho}\,{\bf e}_{\rho}+u_z\,{\bf e}_z.
\ee
In this two-dimensional space, the specular-reflection condition
retains the form of Eq.\ (\ref{speculareqn}),
\be\label{specular2deqn}
{\bf v}_{reflected} = 
{\bf v} - 2\,{\bf u}\,({\bf u}\cdot{\bf v}).
\ee
This is the reason why we can call the motion in the $z$ - $\rho$ plane
a {\em billiard problem}. 

\subsection{Curved ray paths in the centrifugal 
billiard}\label{sec:raylz}
We know that the trajectories between reflections are straight lines,
so that the components of ${\dot{\bf r}}$ in the cartesian coordinate 
frame are constant for each segment. In our new $z$ - $\rho$ frame of 
reference, the $z$-axis is the same as the cartesian one, so that we
still have $v_z={\dot z}={\rm const}$ between reflections in 
Eq.\ (\ref{2dveleqn}). However, the same does {\em not} hold for the 
$\rho$-component of ${\bf v}$. Instead, we obtain from 
Eq.\ (\ref{eneryconseqn})
\be\label{eq:otherenergy}
\dot{\rho}^2+\frac{L_z^2}{\rho^2} = 1-\dot{z}^2={\rm const}.
\ee
If we multiply this by $4\,\rho^2$, it can be written as a
differential equation for $\rho^2$:
\be
\left(\frac{d}{dl}\rho^2\right)^2=4\,\rho^2\,{\dot\rho}^2=
4\,(1-{\dot z}^2)\rho^2 - 4\,L_z^2.
\ee
The solution is that $\rho^2(l)$ describes a shifted parabola,
\be\label{solutioneqn}
\rho^2(l)=\rho^2_0 + 2\,(l-l_i)\,\sqrt{(1-{\dot z}^2)\,\rho_i^2 - L_z^2}
+(l-l_i)^2\,(1-{\dot z}^2),
\ee
where $\rho_i^2$ is the integration constant and specifies the value
of $\rho^2(l_i)$ at the starting point $l_i$ of the ray. Since 
furthermore $z$ is a linear function of $l$ (${\dot z}=const$), we 
can for definiteness fix the initial point is to lie on the focal plane 
and substitute 
\be
l=l(z)=z/{\dot z}
\ee
to find that Eq.\ (\ref{solutioneqn}) describes a curved path 
$\rho(z)$ in the $z$ -$\rho$ plane. 
The curved nature of this trajectory is a direct consequence of the
centrifugal potential $L_z^2/\rho^2$ in 
Eq.\ (\ref{eneryconseqn}), and we would recover straight lines, i.e. 
linear variation of $\rho(l)$ for $L_z=0$. This is why we refer to 
this problem as a {\em centrifugal billiard} \cite{mcbook}. For a 
visual example of how curved traces arise from stright-line 
trajectories, the reader is referred to Fig.\ \ref{fig:chaosl05} (e)
which will be discussed in Section \ref{sec:chaos}.

\begin{figure}[tbp]
    \centering
    \psfig{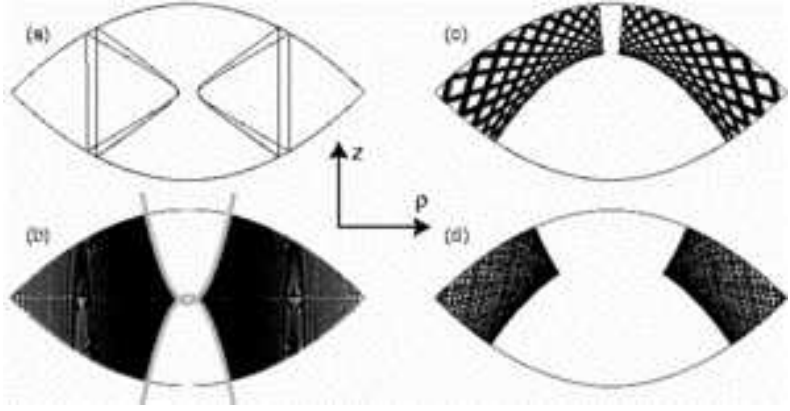}
    \caption{All starting conditions for the rays in (a), (b) and (c) 
    amount to the same angular momentum, $L_{z}=0.1\,f$, but prescribe 
    different angles of incidence with respect to the boundary, and 
    different positions of impact. 
    (a) shows two different orbits in the same plot, both are 
    periodic and symmetric in this projection onto the $z$ - 
    $\rho$ plane, differing only in their starting points. 
    (b) shows a single path, which is quasi-periodic because it does 
    not close on itself even in this projection. Instead, it 
    densely fills a region of space 
    delimited by a caustic whose shape is given by the parabolas 
    (blue). The caustic becomes more 
    asymmetric in (c) and (d), each of which shows a quasiperiodic orbit.
    The angular momentum in (d) is $L_{z}=0.6\,f$, leading to a 
    larger forbidden zone around the $z$ - axis.
        \label{fig:raytrajec}
}
\end{figure}

An example of the ray motion in the special case $L_{z}=0$ is already 
displayed in Fig.\ \ref{fig:n0bowtie}, showing no curved trajectories 
because there the $z$ - $\rho$ plane is indistinguishable from the 
cartesian $z$ - $x$ plane. For $L_{z}\neq 0$, 
curved ray trajectories in the $z$ - $\rho$ plane are shown in 
Fig.\ \ref{fig:raytrajec} for four different initial conditions under 
which the ray is launched. Note that the parameter $L_{z}$ as given in 
the plot has dimensions of length, cf.\ Eq.\ (\ref{eneryconseqn}). 
This reduced two-dimensional problem can be analyzed completely 
without reference to the original three-dimensional ray tracing, with 
$L_{z}$ as a parameter that encapsulates the third degree of freedom 
$\phi$ which has been eliminated. We have, broadly speaking, converted 
to a co-rotating frame of reference (with rotation speed always 
matching the varying angular velocity of the ray), and thus obtained 
a planar problem in which we now look for the classical orbits. The 
simplification is considerable because the three-dimensional ray motion 
in the cavity is rather difficult to visualize, compared to the motion 
in the $z$ - $\rho$ plane.

The two periodic orbits in Fig.\ \ref{fig:raytrajec} (a) exemplify 
this situation: after the trajectory completes one round-trip in the 
$z$ - $\rho$ plane, it returns to its initial position with the 
initial orientation -- but in the original three-dimensional cavity 
there has also been a motion in $\phi$ which does not necessarily 
amount to a full rotation around the $z$ axis. Hence, this periodic 
orbit of the centrifugal billiard is not in general a true periodic 
orbit of the parabolic dome, cf.\ 
Fig.\ \ref{fig:chaosl05} (e). However, we can reverse this statement 
and conclude that any periodic orbit of the three-dimensional problem 
must also be periodic in the $z$ - $\rho$ plane. 
This cautionary remark concerning the 
interpretation of Fig.\ \ref{fig:raytrajec} is relevant if we attempt 
to interpret the actual modes of the original cavity in terms of a 
naive physical optics approach: one might think that a quantized mode 
has to be associated with ray paths that form a closed loop and in 
that way ``reproduce'' themselves. However, a comparison between 
Figs.\ \ref{fig:unsymprofiles} and \ref{fig:raytrajec} reveals
that periodic orbits seem to play no special role for the mode 
structure.

What shapes the modes is not any single periodic ray orbit, but the 
caustics as they appear in Fig.\ \ref{fig:raytrajec} 
(b) - (d).
The spatial distribution of the ray trajectories exhibits a clear 
correspondence with the modal intensities shown in 
Fig.\ \ref{fig:unsymprofiles}, particularly in the shape of 
the caustics. This is most convincing for the two examples in the 
bottom row of Fig.\ \ref{fig:unsymprofiles} where the wavelength is 
shortest: The state at $n=3$, $x=34.16$, $Z=-8.72$ should be compared 
to Fig.\ \ref{fig:raytrajec} (c), and the reflection-symmetric mode 
with $n=4$, $x=73.73$, $Z=0$ finds its counterpart in 
Fig.\ \ref{fig:raytrajec} (d). 

Caustics are immediately generated when we follow a single 
quasiperiodic orbit, but not so for a periodic one. However, periodic 
orbits occur in infinte families which, when plotted together, again 
fill a region of space bounded by a caustic curve. The two members 
of the family shown in Fig.\ \ref{fig:raytrajec} (a) are obtained by 
launching a ray from the focal plane, perpendicular to it, 
differing only in the 
radial distance $\rho$ of the launch.
All other siblings of the examples in Fig.\ \ref{fig:raytrajec} (a)
combined, would create a picture almost identical to the one 
generated by the single quasiperiodic orbit in 
Fig.\ \ref{fig:raytrajec} (b) -- the latter is in fact the result of 
only a slight deviation from the initial conditions chosen in 
Fig.\ \ref{fig:raytrajec} (a), with the result that the orbit {\em 
almost}, but not quite, closes on itself after one round trip, and 
continues to fall short of closing itself after each subsequent round 
trip as well. The conclusion is that from the point of view of the 
caustic structure in our system, there is no qualitative difference 
between periodic and quasiperiodic orbits. 

The fact that all orbits can be characterized by a particular caustic 
which they touch, and that moreover all periodic orbits come in 
infinite families, is a general 
property of {\em integrable} Hamiltonian systems, to which the 
special centrifugal billiard defined here belongs. That the 
paraboloid billiard is integrable, can already be concluded from the 
existence of a separation ansatz for the wave equation, which we 
discussed in Section \ref{sec:waveeq}. However, we have not yet 
completed our program of connecting the ray and wave approaches, and 
in particular we have not addressed the question of how to determine 
quantitatively the type of ray trajectories that correspond to a given 
mode. So far, the correspondence was established by visual inspection 
alone. The quantitative connection is obtained by comparing
the ray patterns of Fig.\ \ref{fig:raytrajec} with the 
effective potential $V$ of Eq.\ (\ref{eq:potential}). We shall see 
that for an integrable system, we can in fact uniquely connect a 
particular caustic with a given mode.

As a final remark concerning the periodic orbits in this integrable 
system, it is worth comparing the patterns of 
Fig.\ \ref{fig:raytrajec} (a) and especially Fig.\ (\ref{fig:n0bowtie})
with the ``bowtie laser'' of Ref.\ \cite{gmachl}. There, a 
semiconductor cavity was designed in such a way as to obtain lasing 
from a bowtie-shaped mode with highly desirable properties, foremost 
among them its focussing action in the center of the cavity. The 
focussing patterns of Fig.\ (\ref{fig:n0bowtie}) are very similar, but 
the main difference is that in our case these orbits occur in families 
whose members can cross the $z=0$ plane with all possible axial 
displacements $\rho$. In the semiconductor cavity, most rays move 
on chaotic trajectories, and only a small range 
of initial conditions for the rays lead to a stable bowtie pattern, leading 
to modes which are strongly concentrated near a unique bowtie path, 
and hence even less spread out in space than the examples shown in 
Fig.\ \ref{fig:symprofilesn1}. This leads us to anticipate that the 
beneficial properties of the $n=1$ modes found for our integrable 
system can in fact be enhanced if we allow for the possibility of 
chaos in the ray dynamics. 

\subsection{Connection with the effective potential in parabolic 
coordinates}
The classical turning 
points for the two degrees of freedom $u$ and $v$ in the potential of
Eq.\ (\ref{eq:potential}) determine the parabolas which describe the 
caustics in Fig.\ \ref{fig:raytrajec}. 
We notice that the caustics (and also the quasiperiodic rays 
that generate the caustics we show) have a well-defined distance of closest 
approach $\rho_{0}$ with respect to the $z$ axis, given by the corner 
at which the two bounding parabolas meet. Describing this in 
parabolic coordinates, we find that $\rho_{0}$ is approached if both 
$\xi$ and $\eta$ simultaneously reach their inner turning points. 
Expressing this condition in terms of Eqs.\ (\ref{eq:uturnpoint}) and 
(\ref{eq:vturnpoint}), we obtain the simple {\em semiclassical} 
relation 
\be\label{eq:rmin}
\rho_{0}=\frac{n}{k}.
\ee
Here, we have used the coordinate transformation 
$\rho=\sqrt{\xi\,\eta}$, cf.\ Eq.\ (\ref{eq:revtransf}), and the 
definition of the rescaled variables, Eq.\ (\ref{eq:uresc}). 

The distance of closest approach for individual {\em periodic} orbits 
is not given by this expression, but the minimal $\rho$ over the 
whole {\em family} of such orbits does follow this law. 
The caustics in  Figs.\ \ref{fig:raytrajec} (b-d) exhibit 
cusp singularities at $\rho_{0}$ because in that extreme point the 
$z$ motion has 
zero velocity: it is clear from Eq.\ (\ref{eneryconseqn}) that the 
smallest $\rho$ will be achieved when ${\dot\rho}={\dot z}=0$. But 
from the same equation we immediately obtain that the angular 
momentum then equals the axial distance, and with 
Eq.\ (\ref{eq:rmin}) this reproduces Eq.\ (\ref{eq:lzsemicl}).
We have therefore established that the ray's ``angular momentum'' 
is directly proportional to the modified angular momentum quantum number 
$n$ of the mode under consideration. In the semiclassical limit of 
large $k$, the difference between $n=m\pm 1$ and $m$ becomes 
negligible in this expression, so that we recover the intuitively 
expected proportionality
\be
L_{z}=\rho_{0}\approx \frac{m}{k}.
\ee
This approximation means that we can neglect the effect of 
polarization on the ray-wave correspondence in the semiclassical 
limit -- however, we shall make use of this only later, 
in the ray analysis of Section \ref{sec:chaos}. Since we have 
been interested in states at rather small $k$ and in particular 
$n\geq 1$, we have 
plotted in Fig.\ \ref{fig:raytrajec} only trajectories with $L_{z}\neq 
0$.

A second semiclassical relation follows from Eqs.\ (\ref{eq:uturnpoint}) 
and (\ref{eq:vturnpoint}) if we ask for the value $z_{0}$ of $z$ 
corresponding to the point $u_{0},\,v_{0}$ at which the 
caustics have their singularities. The whole caustic is uniquely 
determined by its singular point at radial distance $\rho_{0}$ and 
height $z_{0}$, cf.\ Fig.\ \ref{fig:raytrajec}. According to 
Eq.\ (\ref{eq:revtransf}), we get
\be\label{eq:esemicl}
z_{0}=\frac{Z}{2k}.
\ee
This identifies the meaning of the separation constant $Z$, 
also quantifying the earlier observation that for $E=0$ both the wave 
and ray patterns are symmetric with respect to the focal plane: in 
that case, the cusp occurs on this mirror plane, as in 
Fig.\ \ref{fig:raytrajec} (b). 

With Eqs.\ (\ref{eq:lzsemicl}) and (\ref{eq:esemicl}), we have 
completed the bridge from the exact wave equation via semiclassical 
WKB quantization to the ray caustics. By specifying the quantized 
$n$, $Z$ and $k$ of a given mode, we uniquely determine a caustic and with 
it a particular family of ray paths. Now we can use additional 
properties of the ray picture to better understand the cavity modes.
This is especially promising in this system because we have seen that 
the semiclassical approximation is extremely accurate here. The 
reason for this somewhat surprising accuracy lies itself in the 
properties of the ray dynamics, but in order to make this clearer we 
need to introduce the concept of a phase space in which the ray 
dynamics can be described.

\subsection{Families of rays and Poincar{\'e} sections}
\begin{figure}[tbp]
            \centering
        \psfig{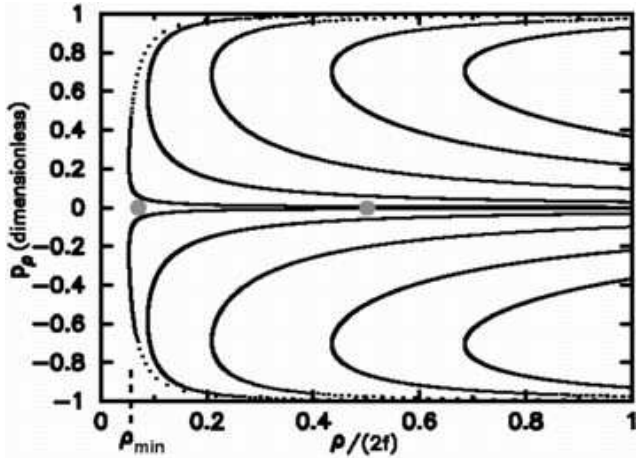}
        \caption{Poincar{\'e} surface of section of the ray dynamics 
        for $L_{z}=0.1\,f$ as in Fig.\ \ref{fig:raytrajec} (a-c). The 
        minimum distance $\rho_{0}$ from the $z$ axis, given by 
        Eq.\ (\ref{eq:rmin}), is indicated by 
        the dashed line. All quasiperiodic trajectories fill smooth 
        curves extending between some minimum 
	$\rho_{\rm min}\geq\rho_{0}$ and $\rho=2\,f$. The two 
        gray dots are the crossings of the focal plane generated 
	by a periodic orbit as shown in Fig.\ \ref{fig:raytrajec} (a).
	The plot uses $2\,f$ as the length unit.
	        \label{fig:sosl05}
}
    \end{figure}

A phase-space description is often used in classical mechanics because 
it carries more information about the possible trajectories than mere 
real-space diagrams. This approach has recently been applied to the 
analysis of ray dynamics in optical cavities as well 
\cite{optlett,nature}, with the goal of providing insights that are 
not revealed by ray tracing in real space. In particular for the 
treatment of non-integrable resonator geometries, it has proved 
valuable to represent the phase space of the rays in terms of 
Poincar{\'e} Surfaces of Section (SOS). 
For our purposes, the following SOS will be chosen:

It is easy to convince ourselves by recalling 
Fig.\ \ref{fig:doubleparab} that any ray trajectory in the cavity has 
to encounter the focal plane infinitely many times as it propagates. 
However, the radial distance of these crossings, as well as the value 
of ${\dot \rho}$ may vary from one crossing of this plane to the 
next. Now we can consider
\be
\rho~~\mbox{and}~~p_{{\rho}}\equiv{\dot\rho}
\ee
as a pair of canonically conjugate position and momentum variables, 
and attempt to image the subset of phase space spanned by them. In 
order to do that, we launch a ray trajectory and follow it for many 
crossings of $z=0$, each time recording the instantaneous values of 
$\rho,\,p_{{\rho}}$ as a point in a two-dimensional graph. The result 
is shown in Fig.\ \ref{fig:sosl05}. A typical trajectory is -- as 
mentioned above -- quasiperiodic, and in the SOS generates a 
dense set of points that all lie on a smooth curve. Several 
trajectories have been followed in this way and are represented in 
Fig.\ \ref{fig:sosl05} by the different individual curves. Each curve 
exhibits some minimal axial distance $\rho_{\rm min}>\rho_{0}$; 
this is a true inequality because quasiperiodic orbits do not 
reach their point of  closest approach to the $z$ 
axis precisely on the focal plane. Since the SOS records 
the instantaneous $\rho$ upon crossing the focal plane, the resulting 
curves have their turning points at larger $\rho$. 

The only orbits 
which have their real turning points exactly at the focal plane are 
the periodic orbits. A 
periodic orbit as displayed in Fig.\ \ref{fig:raytrajec} (a) 
generates exactly two discrete points in the SOS, corresponding to the 
two distinct values of the radial distance $\rho$ at which the axis 
$z=0$ is crossed. Both points in the SOS lie at $p_{\rho}=0$ for the 
periodic orbit, as can be verified from the trajectory in the 
$z$ - $\rho$ plane which always crosses the $z$ axis perpendicularly.  
The quasiperiodic trajectory of Fig.\ \ref{fig:raytrajec} (b)
corresponds to the leftmost curve in 
Fig.\ \ref{fig:sosl05}, which has its turning point almost at $\rho_{0}$ 
in the SOS.
The caustic is almost on the focal plane but still offset from it by 
an amount that is not discernible in Fig.\ \ref{fig:raytrajec} (b). 

The distinction between the periodic orbit and its closely 
neighboring quasiperiodic relative in the SOS of 
Fig.\ \ref{fig:sosl05} is appreciable -- a pair of points generated 
by the 
former, versus a one-dimensional curve for the latter. But exactly on 
the line $p_{\rho}=0$, there exists an infinite number of other pairs 
of points, belonging to the periodic orbits of the same family. 

The SOS in these coordinates allows us to see directly in which 
places the focal plane comes into contact with the rays under 
consideration. This is a central piece of information when it comes 
to estimating the focussing at this plane where the quantum well is 
assumed to be. The forbidden regions around the $z$ axis induced by 
the angular momentum barrier show up as inaccessible portions of the 
SOS toward small $\rho$. 

\subsection{Accuracy of the semiclassical approximation}
We can also comment on the striking accuracy of the semiclassical 
approach in this system. The Poincar{\'e} section shows that almost 
all trajectories (with the exception of the periodic paths) generate 
curves with the same topology: they begin and end at $\rho=2\,f$, with 
one turning point inbetween. There are other integrable systems for 
which the Poincar{\'e} section has a more complicated structure, one 
closely related example being the ellipsoidal cavity\cite{wiersig3d} 
or its two-dimensional counterpart, the ellipse 
billiard\cite{wiersig2d}. 
In that case, the phase space consists of two components in which the 
topology of the trajectories is different: One type of motion 
consists of rays circulating around the perimeter as so-called 
whispering-gallery orbits, the other is a bouncing-ball oscillation 
across the short diameter\cite{keller}. There is a division between 
these two types of trajectories, similar to that between oscillation 
and rotation in a pendulum -- called the separatrix. The WKB 
approximation or its higher-dimensional generalization, named after 
Einstein, Brillouin and Keller (EBK), cannot be applied without 
severe corrections in the vicinity of such a separatrix in phase 
space \cite{arvieu,wiersig3d,wiersig2d}. In our case, this breakdown 
never occurs, and semiclassical results are thus of high accuracy. 
Being a conic section, the parabola can of course be considered as 
a limiting case of the ellipse, with one of its foci moved to 
infinity. This leaves no possibility for bouncing-ball 
trajectories, which leads to the absence of a separatrix. 

Finally, it is worth asking why the sharp corners at the 
intersection between the paraboloid and the focal plane do not cause 
any corrections to our semiclassical treatment, even though the 
surface curvature at these points is clearly much shorter than the 
wavelength. It is known that in such cases {\em diffraction} can occur 
which makes it impossible to explain the mode structure purely based 
on classical orbits\cite{richensberry,sieber}. However, this 
phenomenon is absent for certain special angles subtended by the 
corners. One of these ``benign'' angles is precisely the $90^{\circ}$ 
angle we encounter at the corners of the double paraboloid, 
cf.\ Fig.\ \ref{fig:doubleparab}. When the confocal condition is 
violated so that deviations from a right angle occur at the corners, 
we have to expect diffractive corrections to the semiclassical 
analysis, resulting from classical rays that hit the corners and are 
reflected in an arbitrary direction because the law of specular 
reflection is undefined in that instance. Fortunately, we shall see in 
Section \ref{sec:chaos} that such orbits are far removed from the 
regions of phase space where we expect the important focussing modes 
to lie. 

In this section, we have discussed how the ray dynamics develops 
caustic structure, and how the latter can be represented with the 
help of the Poincar{\'e} section. We have also observed that the 
high-intensity regions in the wave solutions correspond to the ray 
caustics, because there the density of rays is high - in fact 
divergent if we recall the discussion of the classical turning points 
in the effective potential below Eq.\ (\ref{eq:uturnpoint}). Therefore, 
even in situations where we cannot obtain the wave solutions easily, 
their possible intensity distribution can be inferred by investigating 
the ray dynamics first. This will now be carried out for a 
cavity that deviates from the ideal model shape. 

\section{The non-confocal double paraboloid}
\label{sec:chaos}
Having obtained an overview of the types of ray motion that can be 
encounterd in the parabolic dome, and established the connection to 
the mode structure of the full vectorial wave equations via the 
short-wavelength approximation, we now want to introduce a model 
cavity for which the wave solutions cannot be obtained by separation 
of variables. 
The variety of possible deviations from the ideal model geometry of 
Fig.\ \ref{fig:doubleparab} is enormous, so we have to restrict 
attention to certain special distortions that can be expected to be 
generic in some sense. 

\subsection{The model deformation}
The distortions we choose are obtained by 
pulling the two intersecting paraboloids in 
Fig.\ \ref{fig:doubleparab} apart or pushing them together along the 
$z$ axis by an amount $2\,\epsilon$. Specifically, in spherical 
coordinates as a function of polar angle $\theta$, the shape is given 
by 
\be
r(\theta)=\frac{2\,f}{1+\cos\theta}+\frac{2\,\epsilon}{1+\sqrt{1+\epsilon\,
(1-\cos^2\theta)}}.
\ee
The respective foci, which coincide in the integrable model, 
then move off the $\rho$ axis. This non-confocal arrangement 
of the parabolic walls can be viewed as a model for 
fabrication-induced deviations from the ideal cavity shape -- where 
the dome could be slightly too flat or too pointed. It can also be 
interpreted in a different way, taking into account the possibility 
that the boundary condition at the base of the dome is not exactly 
given by Eq.\ (\ref{bdry-plane}), if some penetration of the field 
through the dielectric mirror on the quantum well is taken into 
account. This is of course a realistic expectation, and its effect on 
the wave solutions would be that the TE electric field no longer 
needs to be strictly symmetric under reflection at the focal 
plane. If one maintains that the dome has indeed been fabricated with 
its base in the focal plane, this ``soft'' boundary condition on 
the mirror can be modeled by assuming that our solutions should 
correspond to waves reflected at a plane removed from the dielectric 
interface by some amount $\epsilon$. 

Therefore, the non-confocal double paraboloid is a way of taking 
into account the cumulative effects of 
fabrication uncertainty and soft boundary conditions at the 
dielectric mirror with a single model parameter $\epsilon$, denoting half 
the distance between the foci of the top and bottom parabolic wall in 
the unfolded cavity. One could think that a perturbation theory in 
$\epsilon$ could allow us to use the solutions obtained so far and 
smoothly extend them to the non-confocal situation. This is the 
traditional approach in physics and it is the reason why only simple, 
integrable systems are treated in textbooks on quantum mechanics or 
classical mechanics alike. However, perturbation approaches become 
tedious and even impossible for wave equations whose short-wavelength 
limit (i.e. ray picture) exhibits {\em chaotic dynamics}. The 
difficulties that arise can already be seen without introducing chaos, 
if we try to obtain the wave functions of an ellpsoid-shaped resonator 
as a perturbative expansion starting from the eigenfunctions of a 
spherical cavity. This poses no problems as long as one is interested 
only in modes of the ellipsoid whose topology is analogous to that 
found in the circle \cite{landaulif}. However, as mentioned earlier, 
the ellipsoid exhibits separatrix structure in phase space because there 
exists a type of motion that the sphere does not possess: the 
bouncing-ball trajectories.

Analogous {\em nonperturbative} effects arise in the present model, 
because the distortion can lead to new types of trajectories that are 
not present in the confocal cavity, in a process known as 
bifurcation \cite{reichl,brack}. 
The first consequence of the deformation $\epsilon$ is that 
the infinite families of periodic orbits break up, leaving only 
a distrete number of periodic orbits of the same topology, 
which can be divided in an equal number of stable and unstable paths. 
Stable paths have the property that rays with slightly different 
initial conditions remain close to the given periodic path for all 
times, while unstable periodic orbits are surrounded in their 
immediate neighborhood by chaos -- trajectories deviate from such a 
periodic orbit at an exponential rate if the initial condition is only 
infinitesimally varied. For more quantitative statements and further 
background on the transition to chaos, the reader is referred to the 
literature \cite{reichl,gutzwiller,mcbook}. 

\subsection{Unstable and stable ray motion in the deformed cavity}
The Poincar{\'e} section is very suitable as a diagnostic tool to 
identify this process of emerging chaos on on hand, and the 
stabilization of certain periodic orbits on the other hand. 
    \begin{figure}[tbp]
            \centering
        \psfig{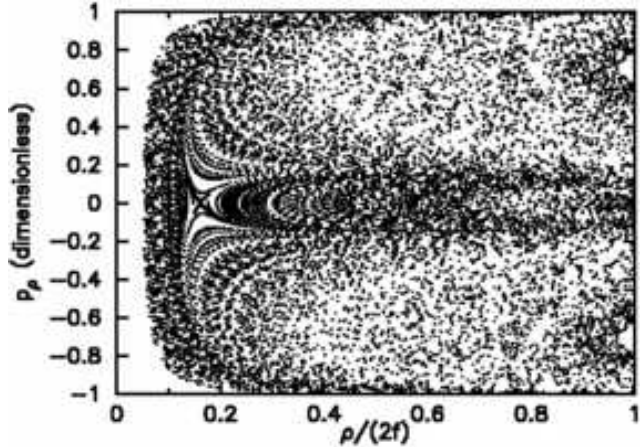}
        \caption{Surface of section at $L_{z}=0.1\,f$ of a non-confocal 
        double-paraboloid, with foci pulled apart by $\epsilon=0.02\,f$. 
	This destabilizes the cavity, leading to chaotic ray dynamics which 
	generates an irregular cloud of points filling almost the whole 
	region that is accessible for this $L_{z}$. A special point is 
	encountered on the line $p_{\rho}=0$ where the 
	irregularity gives way 
	to a confluence of hyperbolic traces whose vertices are centered on
        a single, unstable periodic orbit. The spatial pattern of this new 
	periodic orbit is shown in Fig.\ \ref{fig:chaosl05} (a).
	        \label{fig:sosl05e01}
}
    \end{figure}
This is illustrated in Fig.\ \ref{fig:sosl05e01}. The perturbation 
consists of pulling the foci of the walls apart by $\epsilon=0.02\,f$ 
along the $z$ - axis. Since this preserves the axial symmetry of the 
cavity, $L_{z}$ is still a conserved quantity -- the arguments of 
Section \ref{sec:raylz} rely on no other symmetries of the problem. 
We chose $L_{z}=0.1\,f$ in the plot. 
The small distortion of one percent is already sufficient to change 
the phase space portrait significantly, compared to 
Fig.\ \ref{fig:sosl05}). The unstable periodic orbit appearing 
prominently in Fig.\ \ref{fig:sosl05e01}) as a so-called hyperbolic 
point, is shown in its spatial pattern in Fig.\ \ref{fig:chaosl05} 
(a). It is a self-retracing periodic orbit because it reflects from 
the boundary at normal incidence (in the $z$ - $\rho$ plane). 

The effects that chaos can have on the ray motion are illustrated in 
Fig.\ \ref{fig:chaosl05} (b). Shown there is a single ray trajectory 
which superficially has some similarity to 
Fig.\ \ref{fig:raytrajec} (b). However, the path does not trace out a 
well-defined caustic in Fig.\ \ref{fig:chaosl05} (b). What looks like 
a caustic here is in fact better described as two caustics of the 
type in Fig.\ \ref{fig:raytrajec}, arranged almost symmetrically 
with respect to the focal plane. Note in 
particular the symmetric occurrence of cusps both below and above the 
line $z=0$. Recall that in the integrable case the position 
$\rho_{0}$, $z_{0}$ of 
the caustic singularity is uniquely given by the turning points 
$\xi_{0}$ and $\eta_{0}$ (or equivalently 
$u_{0}$, $v_{0}$), in the effective potential. Reversing the sign 
of the cusp coordinate $z_{0}$ corresponds to 
{\em exchanging} the role of $\xi$ 
and $\eta$. The significance of Fig.\ \ref{fig:chaosl05} (b) is 
therefore that the degrees of freedom $\xi$ and $\eta$ are no longer 
decoupled, because during a single ray trajectory both the cusps at 
$z_{0}$ and $-z_{0}$ are reached. 
By virtue of Eq.\ (\ref{eq:esemicl}), the quantity $Z$ is thus 
not conserved anymore. A trajectory is able to exhibit multiple 
points of closest approach to the $z$ axis and is not strictly 
guided by caustics. 
    \begin{figure}[tbp]
            \centering
        \psfig{file=fig13.epsf,width=4cm}
        \caption{Trajectories in the non-confocal cavity. (a) shows 
        the unstable periodic orbit arising at $\epsilon=0.02\,f$, 
        $L_{z}=0.1\,f$, cf.\ Fig.\ \ref{fig:sosl05e01}. For the 
        same parameter, a chaotic trajectory is seen in (b). 
	Oscillatory motion around stable periodic orbits occurs in (c) and 
	(d), where $\epsilon=-0.02\,f$ and $L_{z}=0.1\,f$ as in 
	 the SOS of Fig.\ \ref{fig:sosl05e-01}. The patterns of type
	 (a) and (c) derive from the periodic motion of 
	 Fig.\ \ref{fig:raytrajec} (a) as a result of the shape 
	 perturbation. In real 
	 three-dimensional cartesian space, (e) shows the straight-line ray 
	 motion (arrows) giving rise to the curved ``envelope'' surface
whose 
	 cross section we see in (a). 
	        \label{fig:chaosl05}
}
    \end{figure}

Under these circumstances, it is not clear what to expect for the mode 
structure of the cavity because we lose the possibility of labeling 
the eigenstates by a complete set of quantum numbers. This does not 
imply there are no modes associated with chaotic rays, but one 
requires additional techniques to perform a semiclassical quantization
\cite{gutzwiller,child}. The destruction of the conserved quantity $Z$ 
means that there is one less constraint which the ray trajectories 
have to satisfy; this allows them to fill two-dimensional areas 
instead of one-dimensional curves in the SOS. Since the SOS gives us 
a picture of how the rays intersect the plane $z=0$, chaotic rays can 
be seen to show less concentrated overlap with that plane. 
We anticipate that the presence 
of true caustics is required to create the best focussing action. 
With this hypothesis, the goal must be to identify ray orbits that 
exhibit caustics. This occurs in the vicinity of stable periodic 
orbits, due to the fact that perturbed trajectories execute an 
oscillatory and in general quasi-periodic motion around such stable 
orbits. In Fig.\ \ref{fig:sosl05e01}, however, 
no stable periodic orbits can be identified, telling us that for the 
deformation chosen ther, no stable modes with $L_{z}=0.1\,f$ should 
exist. 

    \begin{figure}[tbp]
            \centering
        \psfig{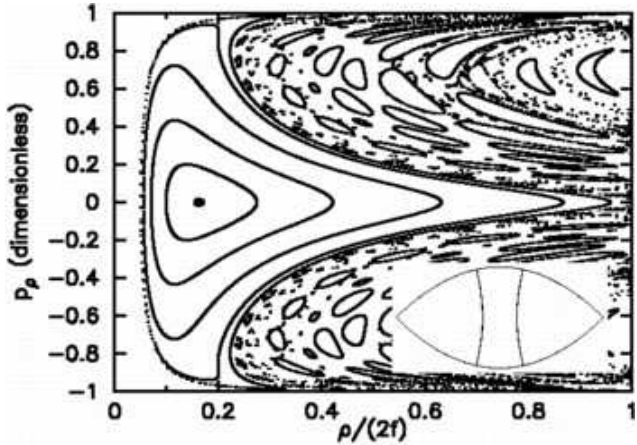}
        \caption{Surface of section at $L_{z}=0.1\,f$ with 
	$\epsilon=-0.02\,f$. The cavity develops a multitude of stable 
	periodic orbits surrounded by elliptical islands. The most prominent
	island of stability is centerd on the the line $p_{\rho}=0$ where a 
	small circle indicates the location of the corresponding periodic 
	orbit around which other trajectories can oscillate. Shown in the 
	inset is the central stable periodic orbit. The next 
	innermost closed line in the SOS belongs to the trajectory shown in 
	Fig.\ \ref{fig:chaosl05} (c).
	        \label{fig:sosl05e-01}
}
    \end{figure}
    The situation changes if we 
consider Fig.\ \ref{fig:sosl05e-01}, in which $L_{z}$ is the same 
but the sign of the non-confocal displacement $\epsilon$ is reversed. 
The walls of the double paraboloid are hence pushed together instead 
of being pulled apart. The resulting phase space structure in the SOS 
differs markedly from Fig.\ \ref{fig:sosl05e01}: many trajectories 
trace out one-dimensional curves in the SOS which organize as closed 
loops, forming island chains that proliferate with various sizes. All 
these islands are centered around stable periodic orbits -- the 
biggest island of stability lies symmetrically around the line $z=0$ 
and corresponds to oscillatory motion of the type shown in 
Fig.\ \ref{fig:chaosl05} (c). The center of the island is in fact 
formed by a periodic orbit similar to Fig.\ \ref{fig:chaosl05} (a) -- 
the only difference being, that small perturbations of its initial 
conditions do not lead to chaos as in Fig.\ \ref{fig:sosl05e01}, but 
to the motion of Fig.\ \ref{fig:chaosl05} (c). 

Another oscillatory 
ray path centered at a stable periodic orbit is shown 
in Fig.\ \ref{fig:chaosl05} (d). The pattern should be compared to 
Fig.\ \ref{fig:raytrajec} (c) which has the same $L_{z}$. The 
similarity is apparent, except for the fact that the path in 
Fig.\ \ref{fig:raytrajec} (c) will eventually fill the remaining gaps 
in that plot, if one follows it longer. The path in 
Fig.\ \ref{fig:chaosl05} (d), on the other hand, is truly 
restricted to the vicinity of a self-retracing orbit which reverses 
its propagation direction at one end due to perpendicular reflection 
at the wall, and at the other end by running up the centrifugal 
barrier perpendicular to the $z$ axis. 

All islands of stability in Fig.\ \ref{fig:sosl05e-01} generate their 
own caustics, which are topologically different from the ones in the 
integrable system. The caustic created by the orbit in 
Fig.\ \ref{fig:chaosl05} (c) is simply the boundary of the 
regions into which the ray never penetrates. The difference between 
the absence and presence of caustics 
in Figs.\ \ref{fig:chaosl05} (b) and (c) is not easily 
appreciated if we consider only the real-space plots. Here, the 
usefulness of the Poincar{\'e} section as a diagnostic tool is again 
to be noted -- showing two-dimensional clouds of points versus 
one-dimensional curves, respectively, for trajectories without and 
with caustics. 
\begin{figure}[tbp]
            \centering
        \psfig{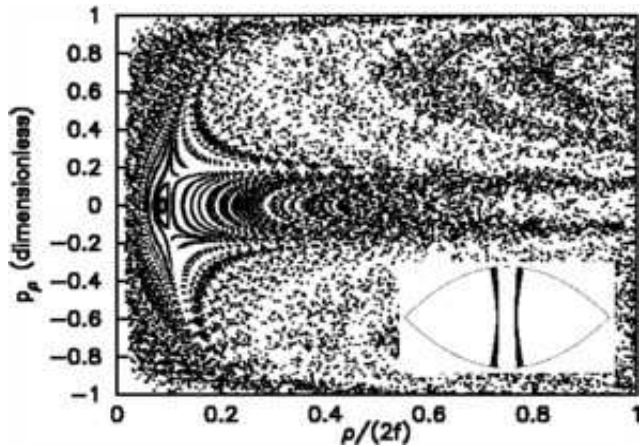}
        \caption{For the same deformation $\epsilon=0.02\,f$ as in 
	Fig.\ \ref{fig:sosl05e01}, this 
	surface of section at the smaller angular momentum $L_{z}=0.03\,f$ 
	shows that a stable orbit exists in addition to the unstable 
	hyperbolic one. This is indicated by the elliptic 
	(lens-shaped) island structure. 
	The hyperbolic point is located to the right of the island. The ray 
	pattern near the stable periodic orbit is shown in the inset. The 
	small corresponding mode volume is apparent. 
	       \label{fig:sosl015e01}}
 
    \end{figure}

It follows from the preceding discussion that a negative $\epsilon$ 
leaves us with a cavity that is in many respects similar to the 
unperturbed double paraboloid, cf. the ray pattern of 
Fig.\ \ref{fig:chaosl05} (d). However, the qualitative and important 
difference is that some periodic orbits are now {\em more stabilized} 
than at $\epsilon=0$. In particular, there are simple ray 
bundles such as Fig.\ \ref{fig:chaosl05} (c) that promise reasonable 
focusing close to the center of the unfolded cavity. The physical 
explanation for the general stabilizing effect that we achieved by 
moving the paraboloids closer together lies in the well-known fact 
that a two-mirror resonator configuration has a a focusing action 
when the mirrors are separated less than the sum of their radii of 
curvature. Conversely, mirrors that are further apart than this 
criterium act in a defocusing way. This is consistent with the 
observation of a large chaotic domain in Fig.\ \ref{fig:sosl05e01}. 

These simple arguments, and the chaotic picture of 
Fig.\ \ref{fig:sosl05e01}, seem to suggest that stable ray motion is 
not to be expected in the supposedly defocusing configuration 
with $\epsilon=0.02\,f$. However, 
when applying the standard criteria for focusing and defocusing 
resonator geometries, we have to bear in mind that we are dealing with 
a centrifugal billiard whose ray trajectories are curved. The effect 
of the centrifugal barrier is to push the regions of allowed ray 
motion outwards until only a small patch surrounding the equatorial 
corners of the cavity is accessible. At large $L_{z}$ the motion is 
then so confined that chaos does not develop. This is just the 
whispering-gallery phenomenon\cite{mcbook}. On the other hand, at 
$L_{z}=0.1\,f$ we certainly found chaos with no remaining 
islands of stability. Small $L_{z}$ are what we must be interested in 
if concentration near the focal points is to be achieved.

In view of this, it is all the more surprising that the same cavity 
does in fact support stable orbits at even {\em smaller} angular
momenta than in  Fig.\ \ref{fig:sosl05e01}. This is 
shown in  Fig.\ \ref{fig:sosl015e01} for 
$L_{z}=0.03\,f$. The periodic orbit responsible for the single 
stable island in that SOS is again almost identical to the one shown 
in Fig.\ \ref{fig:chaosl05} (a), and its oscillatory neighborhood is 
analogous to Fig.\ \ref{fig:chaosl05} (c); the inset of 
Fig.\ \ref{fig:sosl015e01} shows this similarity. 
This stable orbit exists 
only at sufficiently small $L_{z}$; its associated island in the SOS
shrinks to a point when $L_{z}\approx 0.038\,f$. 
The conclusion is that {\em both} the nominally focusing and 
defocusing configurations $\epsilon=\pm 0.02\,f$ permit the 
formation of ray bundles with a spatial distribution as in 
Fig.\ \ref{fig:chaosl05} (c), and hence the stable modes associated 
with this pattern should be robust. This is also confirmed by 
analogous Poincar{\'e} sections for larger displacements of the foci. 
At larger $|\epsilon|$, the motion of type Fig.\ \ref{fig:chaosl05} (c) 
and the inset of Fig.\ \ref{fig:sosl015e01} 
is in fact stabilized further -- for both directions of displacements 
{\em alike}. 

The modes corresponding to this particular ray pattern are closely 
related to the fundamental s-waves we discussed in 
Section \ref{sec:fundawave}, because both arise from ray bundles in 
the immediate vicinity of the {\em shortest} periodic orbits in the 
cavity. For $\epsilon=0$ this was the family of paths in 
Fig.\ \ref{fig:raytrajec} (a), members of which can be smoothly 
deformed into Fig.\ \ref{fig:chaosl05} (a) without changing the 
topology -- i.e., the number and sequence of reflections and turning 
points. We shall therefore call all these orbits the {\em fundamental 
orbits} of the cavity. 
The mode spacing of the corresponding eigenstates should be 
comparable as well for the perturbed and unperturbed case. However, 
we have to defer a detailed analysis of the wave solutions and their 
semiclassical correspondence to a future paper. Here, the goal has 
been to introduce the ray dynamics and its phase space as the 
backbone on which the mode structure is built. 

Assuming that the deformation is $\epsilon=0.02\,f$, we have the 
peculiar situation that the fundamental orbit is unstable if 
$L_{z}>0.038\,f$, cf.\ Fig.\ \ref{fig:sosl05e01}. Therefore, the most 
desirable modes will be those with smaller $L_{z}$. According to 
Eq.\ (\ref{eq:lzsemicl}), we have to choose modes with low 
$n$ and high $k$ to achieve this. For the experimental cavity 
we have $k\,f\approx 14\pi$. Taking $n=1$ as in 
Section \ref{sec:fundawave}, we arrive at the semiclassical value
\be
L_{z}=\frac{f}{14\,\pi}\approx 0.023\,f
\ee
which is close to the situation depicted in Fig.\ \ref{fig:sosl015e01}. 
The difference in the SOS is insignificant. We have no 
accurate way of determining the actual value of $\epsilon$ most 
closely describing the real structure, but these considerations give 
us considerable confidence that modes with a spatial pattern as in 
Fig.\ \ref{fig:chaosl05} (c)  or Fig.\ \ref{fig:sosl015e01}
will be found in the cavity, because the 
relevant $L_{z}$ estimated above is in a range where this fundamental 
orbit is stable -- {\em irrespective} of the sign of $\epsilon$ and 
moreover largely independent of its magnitude. 

\section{Bragg mirror as an escape window in phase space}
\label{sec:rayescape}
The internal ray dynamics of the dome resonator has up to this point 
been evaluated under the assumption that the cavity is a perfect 
resonator. There are two physical mechanisms that invalidate this 
viewpoint: absorption in the gold mirror and transmission through the 
Bragg grating. The trade-off between the comparatively large 
absorption of a metal on the one hand and its ability to reflect 
omnidirectionally have been discussed in Ref.\ \cite{metalbragg}. In 
our context, metallic absorption will always degrade the Q factor 
because the gold layer provides only an estimated $95$\% 
reflectivity \cite{metalbragg}. 
However, the reflectivity of the Bragg mirror can be 
significantly {\em lower} for certain modes and in that case 
constitutes the dominant mechanism for Q-spoiling. The variable that 
determines the reflectivity of the Bragg mirror (at the fixed 
operating frequency) is the {\em angle of incidence} $\chi$ with 
respect to the $z$-axis. For the purposes 
of a qualitative analysis, we assume that the Bragg reflectivity is 
unity for $\chi< 22^{\circ}\equiv\chi_{c}$ but drops to $\approx20$\% 
outside this cone of incidence \cite{metalbragg}. 
In other words, $\chi_{c}$ is the boundary between 
absorption-dominated and leakage-dominated Q factors. A second window 
of high reflectivity opens for rays at very oblique incidence on the 
grating surface, more specifically for $\chi>60^{\circ}$. This second 
window will be discussed further below.

The ray picture allows us to use this rough transmission criterion as 
a guide in 
order to separate long-lived cavity modes from short-lived ones. The 
angle $\chi$ between $z$-axis and a trajectory is, according to 
Eq.\ (\ref{eq:unitvector}), given by 
\be
\cos\chi={\dot{\bf r}}\cdot{\bf e}_{z}={\dot z},
\ee
so that 
\be
1-{\dot z}^2=\sin^2\chi.
\ee
One can substitute this as the righthand side of 
Eq.\ (\ref{eq:otherenergy}) and obtains an equation for a curve in 
the plane ${\dot\rho},\,\rho$ (${\dot\rho}=p_{\rho}$) 
spanning the Poincar{\'e} section:
\be
\vert p_{\rho}\vert=\sqrt{\sin^2\chi-\frac{L_{z}^2}{\rho^2}}.
\ee
Using the critical value of $\chi_{c}$ in this equation specifies the 
escape condition in the Poincar{\'e} section: the Bragg mirror 
becomes ineffective when 
\be\label{eq:reflectgood}
\vert p_{\rho}\vert>\sqrt{\sin^2\chi_{c}-\frac{L_{z}^2}{\rho^2}}.
\ee

\begin{figure}[tbp]
            \centering
        \psfig{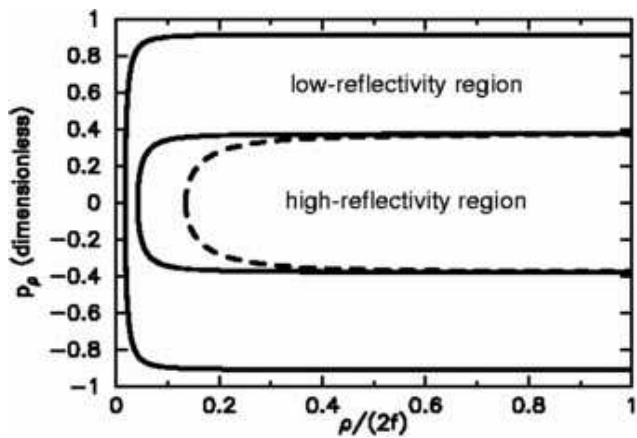}
        \caption{
Assuming that the Bragg mirror stack at the plane $z=0$ yields high 
reflectivity only for waves within $\chi_{c}=22^{\circ}$ from the surface 
normal or for, the regions of high reflectivity in the Poincar{\'e} section 
are bounded by Eq.\ (\ref{eq:reflectgood}). The resulting curves 
bounded by $\vert p_{\rho}<0.374$ are 
shown for $L_{z}=0.03$ (solid line) and $L_{z}=0.1$ (dashed).  
A second high-reflectivity window exists for rays falling between the 
boundary of this plot and the solid line near the boundary. It 
becomes relevant only for the integrable confocal cavity because the 
perturbed shapes have no stable orbits in this second window. 
	        \label{fig:escapecond}
}
    \end{figure}

In order to get a feeling for the type of ray orbits that can remain 
in the cavity under this escape condition, we plot in 
Fig.\ \ref{fig:escapecond} the resulting curves in the surface of 
section for the two different values of $L_{z}$ appearing in 
Figs.\ \ref{fig:sosl05}, \ref{fig:sosl05e01}, \ref{fig:sosl05e-01}
and \ref{fig:sosl015e01}. The plot should be superimposed on these 
plots to decide which parts of the respective phase space falls 
within the high-reflectivity range of the DBR grating. 
Note that the critical lines for ray escape are 
independent of deformation because they rely only on 
Eq.\ (\ref{eq:otherenergy}).

As a result of this comparison, we find first of all that low angular 
momenta are required by the escape criterion, because the phase-space 
area enclosed by the critical curves in Fig.\ \ref{fig:escapecond}
shrinks with increasing $L_{z}$. This is understandable because the 
ray motion in this case has a strong azimuthal component contributing 
to the tilt angle with respect to the $z$-axis. Let us turn our 
attention to the stable periodic orbits arising in the chaotic 
Poincar{\'e} sections. The case 
$L_{z}=0.1\,f$ shown previously for illustrative purposes turns out 
now to be roughly the maximum angular momentum at which the stable 
orbit of Fig.\ \ref{fig:sosl05e-01} is still confined by Bragg 
reflection. The lower angula momentum $L_{z}=0.03\,f$ coming close to 
the estimated value for the s-waves of our experimental cavity, on the 
other hand, places the stable periodic orbit well inside the 
high-reflectivity range of the DBR. For the case of a defocusing 
deformation this is illustrated in Fig.\ \ref{fig:sosl015e01}. The 
peridic point is at $\rho\approx 0.086$. For a focusing deformation 
of the same magnitude, $\epsilon=-0.02$, the periodic point lies at 
$\rho\approx 0.99$. Both values are to the right of the solid line 
in Fig.\ \ref{fig:escapecond}, corresponding to high reflectivity.

For the {\em chaotic} orbits, we observe that they spread out 
over the Poincar{\'e} section in such a way as to yield significant 
overlap with the low-reflectivity regions of 
Fig.\ \ref{fig:escapecond}. This is true for all Poincar{\'e} 
sections shown in this paper. Therefore, we conclude that 
{\em cavity modes 
associated with the chaotic phase space regions are short-lived}, and 
the corresponding broad resonances will not affect the spontaneous 
emission enhancement of the parabolic dome. A quantitative estimate 
of the resonance lifetimes could be obtained by measuring the time 
that a chaotic trajectory spends, on average, in the high-reflectivity 
region without excursions beyond the critical line. However, we shall 
not attempt quantitative predictions at this stage of our 
investigation, and defer it to future work. 

\begin{figure}[tbp]
            \centering
        \psfig{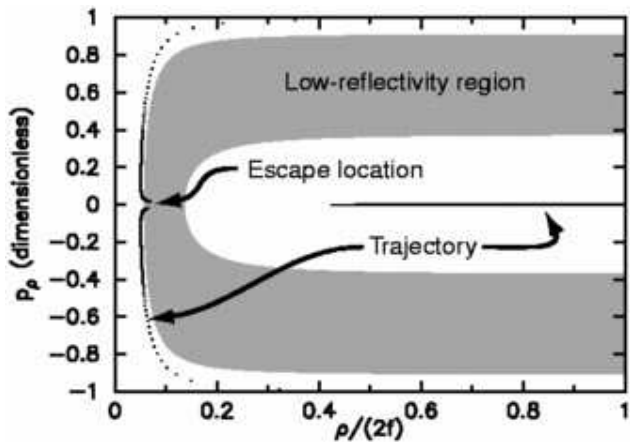}
        \caption{
	The Poincar{\'e} section combined with the escape conditions 
	can be used to extract 
	information about the lifetime and escape locations. This is 
	illustrated here for a single ray orbit (black trace), followed for 
	$500$ crossings of the focal plane. The gray area is the region
which 
	has to be avoided by the ray in order to remain in the cavity. 
	        \label{fig:integrablesc}
}
    \end{figure}

A quantitative analysis would also be necessary to determine the 
modal lifetimes in the marginal case of the 
ideal {\em confocal} cavity. The reason is that the ray picture alone 
does not allow a clear distinction between classically confined and 
unconfined orbits, because the classification according to stable and 
unstable trajectories does not apply in the integrable parabolic 
dome.  All the solid
curves in the Poincar{\'e} section of Fig.\ \ref{fig:sosl05} cross 
into the low-reflectivity region of Fig.\ \ref{fig:escapecond} at 
some point, but the time spent in the high-reflectivity range can be 
very long classically. To illustrate this, we show in 
Fig.\ \ref{fig:integrablesc} a particular ray trajectory for 
$L_{z}=0.1\,f$ in the confocal paraboliod, which for almost $500$ 
crossings of the focal plane remains inside the regions of high 
reflectivity. This time, the second window of high reflectivity close 
to the border of the SOS is important because the ray alternates 
between the low- and high-$\chi$ windows from one crossing of the 
focal plane to the next. The regular nature of this motion makes long 
lifetimes possible because it strictly prevents the ray from 
entering the low-reflectivity region for long times, whereas a 
chaotic orbit would quickly explore this domain in a quasi-random 
way. 

The trajectory shown in Fig.\ \ref{fig:integrablesc} is 
practically identical to the one shown in Fig.\ \ref{fig:raytrajec} (b).
The alternating way of intersecting the focal plane can be understood 
from that figure, or from  Fig.\ \ref{fig:raytrajec} (a) which shows 
periodic orbits closely neighboring the quasiperiodic trajectory of 
plot (b). 
Note that the ray model allows us in addition to predict the spatial 
location where the mode corresponding to this ray bundle will 
preferentially be coupled out through the Bragg mirror. As can be 
seen in Fig.\ \ref{fig:integrablesc}, the low-reflectivity region is 
reached for the first time when, after many reflections, the 
trajectory departs from the immediate neighborhood of the focal 
region, i.e. intersects the focal plane with a $\rho$ that is 
slightly too large. 

The subtle balance of parameters that 
prevents chaos from appearing will, in all experimental realizations, 
be shifted to either the defocusing or the focusing side. 
Therefore, the above ray analysis of the mixed phase spaces for these 
two situations above is our main concern. However, as in the previous 
sections the integrable case is a useful starting point to illustrate 
our strategy. The advantage of the ray approach is that it provides 
fast and intuitive predictions, but further studies are required in 
order to determine how this model succeeds in characterizing the 
cavity quantitatively. 
Paradoxically, we can already conclude that the 
existence of chaos and islands of stability makes it easier to obtain 
results from a ray analysis, because there is a sharper 
separation between long lifetimes for the stable modes discussed above 
and short lifetimes for modes associated with the chaotic portions of 
the SOS.

\section{Conclusion}
In this paper we have examined the modal structure of the electromagnetic
field in a semiconfocal plano-parabolic cavity (or, equivalently, in a
double-paraboloid confocal cavity)
in view of our recent fabrication of semiconductor microcavities
having that geometry.
In order to account for the effects of the inevitable fabrication defects
we also considered the stability of the modes with respect to deformations 
consisting of deviations with respect confocality.
This theoretical analysis was thus motivated by our ongoing experiments
on these structures, and feeds back into this experimental work 
by opening a novel perspective in terms of investigating
the chaotic structure and dynamics of some of the modes of cavity.

Regarding the structure of the modes in the parabolic cavity, we note that
the scalar wave equation is solvable analytically by separation of variables. 
However, the vectorial boundary conditions for the electromagnetic field destroy 
this property, leaving only the cylindrical symmetry. 
Nevertheless, the fundamental series of s-waves (free of azimuthal nodes) in 
a confocal electromagnetic cavity can be solved rigorously.
It has its energy concentrated in a small volume (of order $\lambda^3$) 
around the focal point, even though at the focal point itself
the electric field is zero due to the vectorial nature of the field.
The higher order modes cannot be solved as readily in the full
three-dimensional model, but it is possible to appreciate their features by reducing the
problem to scalar form.  
In these higher order modes,
the energy is concentrated in lobes that surround the focal point but avoid it
because of the centrifugal barrier that arises from the cylindrical symmetry.
Indeed, these modes correspond to non-zero values of the angular momentum
($m\neq 0 $) and for large values of 
$m$ tend towards a type of whispering-gallery modes with intensity 
concentrated in a ring along the focal plane, 
[cf.\ Fig.\ \ref{fig:raytrajec} (d)].

The stability of the modes of the parabolic cavity 
with respect to geometrical deformations can be assessed
by examining the ray trajectories that correspond to each mode.
For a deformation that corresponds to a small deviation from confocality,
chaotic ray patterns emerge. However, we also find stable ray orbits 
concentrated in a small part of the cavity volume. 
Independent of deformation, the most important 
stable orbits being those which in cylinder coordinates $\rho$ and $z$ 
follow the shortest possible periodic trajectory. This general topology 
is the same for a range of deformations (including the ideal confocal 
cavity) and corresponds to a ray 
returning to the same $\rho$ and $z$ after two reflections, missing 
the focal point by a small amount because the field there has to vanish. 
The generic shape of this orbit is represented in 
Fig.\ref{fig:chaosl05}, and its special modification in the confocal 
case with its marginal stability is shown in Fig.\ref{fig:raytrajec}. 
The topological equivalence between the stable orbits of the 
distorted cavity on one hand and of the confocal system on the other 
indicates that the structure of the fundamental
s-wave is stable with respect to deformations.

From the experimental viewpoint, the results of this theoretical analysis indicate
that the cavities already fabricated in our laboratory should possess stable modes
in which the energy is confined in a volume of order $\lambda^3$
in the vicinity of the focal point, in spite of fabrication errors.
The higher order modes, in which the field is concentrated
away from the focal point, in whispering-gallery type configurations,
will be unstable because of the presence of fabrication defects.
At the same time these modes will decay very fast
as they correspond to oblique incidences onto the Bragg mirror,
at angles for which the mirror is no longer reflecting.
Experiments are in progress to characterize the structure and dynamics
of both the stable and unstable modes \cite{next}.
The robust stable modes in which the field is confined in the vicinity of the focal point
should give rise to strong enhancement of the spontaneous emission of a dipole 
(such as a semiconductor quantum well or a semiconductor quantum box)
placed there, and a concomitant lowering of the lasing threshold,
even for our cavities that are of mesoscopic dimensions.
This is because because even in such large cavities, 
whose geometric volume is of the order of a few thousand cubic wavelengths,
the central lobe of the fundamental s-wave (which contains most of the energy) 
has an effective volume of the order of one cubic wavelength. 

These considerations underscore the interest that parabolic microresonators present
by exhibiting quantum electrodynamic effects as well as optical chaos, 
in spite of their relatively large dimensions.
In addition, the mesoscopic cavity dimensions of these structures
are an important practical feature, as they make the fabrication accessible
to existing experimental techniques (such as Focused Ion Beam etching)
while, at the same time, they greatly facilitate the theoretical analysis of these devices
as they permit the use of short-wavelength approximations.

\noindent
{\bf Acknowledgements}:
This work was supported in part
by the European Commission through an ESPRIT-LTR contract
(No. 20029 "ACQUIRE") and a 
TMR Network ("Microlasers and Cavity Quantum Electrodynamics").

\begin{appendix}
\section{Parabolic coordinates}
The parabolic coordinates $\xi, \eta, \phi$ are related to the 
three-dimensional cartesian coordinates according to
\begin{equation}
\left \{ \begin{array}{l}
x = \sqrt{\xi\eta} \cos\phi\\
y = \sqrt{\xi\eta} \sin\phi\\
z = \frac12 (\xi - \eta)
\end{array} \right.
\end{equation}
Or, equivalently,
\begin{equation}
\left \{ \begin{array}{l}
\xi = r+z\\
\eta = r-z\\
\phi = \arctan \frac{y}{x}
\end{array} \right.
\end{equation}
where $r = \sqrt{x^2+y^2+z^2}$ is the spherical radius vector.
With this definition, $\xi$ and $\eta$ have the same dimensions as 
the cartesian coordinates, which is helpful for physical 
considerations. The surfaces $\xi = constant$ are paraboloids by revolution 
about the positive $\hat{z}$-axis having their focal point at the 
origin, while the surfaces $\eta = constant$ are directed along
the negative $\hat{z}$-axis.  The plane $z=0$ corresponds to the 
condition $\xi = \eta$.
In terms of the cylindrical coordinates $\rho=\sqrt{x^2+y^2}, z, \phi$
the parabolic coordinates obey
\begin{equation}\label{eq:revtransfapp}
\left \{ \begin{array}{l}
\rho = \sqrt{\xi\eta}\\
z = \frac12 (\xi - \eta)
\end{array} \right.
\end{equation}
and
\begin{equation}
\left \{ \begin{array}{l}
\hat{\rho} = \frac{1}{\sqrt{\xi + \eta}} 
            \left ( \sqrt\eta \cdot \hat\xi + \sqrt\xi \cdot \hat\eta \right
)\\
\hat{z} = \frac{1}{\sqrt{\xi + \eta}}
            \left ( -\sqrt\xi \cdot \hat\xi + \sqrt\eta \cdot \hat\eta
\right ) 
\end{array} \right.
\end{equation}
In these parabolic cordinates, the electric field
$E = (E_\xi, E_\eta, E_\phi)$ is related to its representation in
cylindrical 
coordinates according to
\begin{equation}
\vec{E} = \left \{
\begin{array}{l}
E_\xi = \sqrt{\frac{\eta}{\xi+\eta}} \frac{i}{\sqrt2}\left ( E_+-E_- \right
) -
\sqrt{\frac{\xi}{\xi+\eta}} E_z
 \\ \\
E_\eta = \sqrt{\frac{\xi}{\xi+\eta}} \frac{i}{\sqrt2}\left ( E_+-E_- \right
) +
\sqrt{\frac{\eta}{\xi+\eta}} E_z \\ \\
E_\phi = \frac{1}{\sqrt2}\left ( E_++E_- \right ) 
\end{array}
\right.
\end{equation} 
\end{appendix}


\begin{thebibliography}{99}
\bibitem{Slusher} 
R.~E.~Slusher and C.~Weisbuch,
Solid~State~Commun. {\bf 92}, 149 (1994)
\bibitem{Bjork}
G.~Bj{\"o}rk, S.~Machida, Y.~Yamamoto, and K.~Igeta, 
Phys. Rev. A {\bf 44}, 669, (1991) 
\bibitem{disk}
S.~L.~McCall, A.~F.~J.~Levi, R.~E.~Slusher, S.~J.~Pearton, and R.~A.~Logan,
Appl. Phys. Lett. {\bf 60}, 289 (1992)
\bibitem{sphere}
L.~Collot, V.~Lef{\`e}vre-Seguin, M.~Brune, J.~M.~Raimond, and S.~Haroche,
Europhys. Lett. {\bf 23}, 327 (1993)
\bibitem{Feld} 
D.~J.~Heinzen, J.~J.~Childs, J.~E.~Thomas, and M.~S.~Feld,
Phys. Rev. Lett.  {\bf 58}, 1320 (1987)
\bibitem{Yamamoto}
F.~M.~Matinaga, A.~Karlsson, S.~Machida, Y.~Yamamoto, T.~Suzuki, Y.~Kadota,
and M.~Ikeda,
Appl. Phys. Lett.  {\bf 62}, 443 (1993)
    \bibitem{next}
    I.~Abram {\em et al.}, in preparation
    \bibitem{fib}
The Focused Ion Beam etching process 
was carried out by Orsay Physics S.A.\ under contract
    \bibitem{laabs}
    H.~Laabs and A.~T.~Friberg, IEEE J.~Quant.~Electron.~{\bf 35}, 
    198 (1999)
    \bibitem{kerker}
    M.~Kerker, {\em The Scattering of Light and Other Electromagnetic 
    Radiation} (Academic Press, New York, 1969)
    \bibitem{siegman}
    A.~E.~Siegman, {\em Lasers} (University Science Books, Mill 
    Valley, California, 1986)
    \bibitem{balianbloch}
    R.~Balian and C.~Bloch, Ann.~Phys.~{\bf 60}, 401 (1970); 
    Ann.~Phys.~{\bf 64}, 271 (1971)
    \bibitem{abramovitz}
    M.~Abramovitz and I.~A.~Stegun, eds., {\em Handbook of 
    Mathematical functions} (Dover Publications, New York, 1972)
    \bibitem{mathews}
    J.~Mathews and R.~L.~Walker,
    {\em Mathematical Methods of Physics}, 
    (Addison-Wesley, Reading, MA, 1970)
    \bibitem{substitution}
The first 
derivative can in principle be removed by another substitution of 
variables, $\rho=-\ln u$. All quantities entering the WKB procedure 
are, however, unchanged by this substitution. Therefore, the following 
discussion retains $u$ as the variable.
    \bibitem{yamamotoslusher}
    Y.\ Yamamoto and R.\ E.\ Slusher, {\it Physics Today} 
    {\bf 46} (6), 66 (1993)  and references therein
    \bibitem{mcbook}
    J.~U.~N{\"o}ckel and A.~D.~Stone, in: {\em Optical Processes in
    Microcavities}, edited by R.~K.~Chang and A.~J.~Campillo (World
    Scientific, Singapore, 1996)
    \bibitem{gmachl}
    C.~Gmachl, F. Capasso, E. E. Narimanov, J. U. N\"{o}ckel,
    A. D. Stone, J. Faist, D. L. Sivco, and A. Y. Cho, Science,
    \textbf{280,} 1556 (1998)
    \bibitem{weyl}  
    H.~Weyl, J.~Reine Angew.~Math.~{\bf 143}, 177 (1913); H.~Weyl, 
    Bull.~Am.~Math.~Soc.~{\bf 56}, 115 (1950)
    \bibitem{keller} 
    J.~B.~Keller and S.~I.~Rubinow, Ann.~Phys.~{\bf 9}, 24
    (1960)
    \bibitem{arvieu}
    R.~Arvieu, F.~Brut, J.~Carbonell and J.~Trouchard, 
    Phys.~Rev.~A {\bf 35}, 2389 (1987)
    \bibitem{wiersig3d}
    P.~H.~Richter, A.~Wittek, M.~P.~Kharlamov and A.~P.~Kharlamov, 
    Z.~Naturforsch.~{\bf 50 a}, 693 (1995);
    H.~Waalkens, J.Wiersig and H.~Dullin, Ann.~Phys.~{\bf 267}, 64 
    (1999)    
    \bibitem{richensberry} P. J. Richens and M. V. Berry, Physica,
\textbf{2D}, 495 (1981)
    \bibitem{sieber} J. B. Keller, J. Opt.\ Soc.\ Am., \textbf{52}, 116
    (1962); M. Sieber, N. Pavloff, and C. Schmit, Phys.\ Rev.\ E,
    \textbf{55}, 2279 (1997)
    \bibitem{wiersig2d}
    H.~Waalkens, J.Wiersig and H.~Dullin, Ann.~Phys.~{\bf 260}, 50 
    (1997)
    \bibitem{optlett} 
    J.~U.~N{\"o}ckel, A.~D.~Stone and R.~K.~Chang, Opt.~Lett.~{\bf 19}, 1693
    (1994)
    \bibitem{nature}
    J.~U.~N{\"o}ckel and A.~D.~Stone, Nature {\bf 385}, 45 (1997)
    \bibitem{landaulif}    
    L.~D.~Landau and E.~M.~Lifshitz, {\em Quantenmechanik}, 
    (Akademie-Verlag Berlin, 1979)
    \bibitem{reichl} L.~E.~ Reichl, 
    {\it The Transition to Chaos}, Springer-Verlag (1992)
\bibitem{brack}
M.\ Brack and R.\ K.\ Bhaduri, {\em Semiclassical Physics}, 
Frontiers in Physics {\bf 96} 
(Addison-Wesley, Reading, USA, 1997)			       
    \bibitem{gutzwiller}
    M.~C.~Gutzwiller, {\em Chaos in Classical and Quantum Mechanics}
    (Springer, New York 1990)
    \bibitem{child}
    M.~S.~Child, {\em Semiclassical Mechanics 
    with Molecular Applications}, 189 (Clarendon Press, Oxford, 1991)
    \bibitem{metalbragg}
    I.~Abram, I.~Robert and R.~Kuszelewicz, IEEE J. Quantum Electronics 
{\bf 34}, 71 (1998)
\end{thebibliography}
\end{document}